# Dynamical system modeling of self-regulated systems undergoing multiple excitations: first order differential equation approach

Denis Mongin (1,2), Adriana Uribe (2), Julien Gateau (3), Baris Gencer (4), Jose Ramon Alvero-Cruz (5) Boris Cheval (1,2), Stéphane Cullati (1,2,6), Delphine S. Courvoisier (1,2)

(1) Quality of Care Division, Geneva University Hospitals, Switzerland

(2) Department of General Internal Medicine, Rehabilitation and Geriatrics, University of Geneva, Switzerland

(3) Galatea Laboratory, EPFL Lausanne, Switzerland

(4) Cardiology Division, Geneva University Hospitals, Switzerland.

(5) Department of Human physiology, histology, pathological anatomy and physical education, Malaga University, Andalucía Tech., Malaga, Spain

(6) Swiss NCCR "Lives: Overcoming Vulnerability: Life Course Perspectives", University of Geneva, Switzerland

Author Note

Corresponding author: Denis Mongin, Department of General Internal Medicine, Rehabilitation and Geriatrics, avenue de Beau-Séjour 26, CH-1211 Geneva. E-mail: denis.mongin@unige.ch.
The code used for this study is given in Annex, and has been released as an R package: the DOREMI package (https://cran.r-project.org/web/packages/doremi/index.html )

## Abstract

This article proposes a dynamical system modeling approach for the analysis of longitudinal data of self-regulated systems experiencing multiple excitations. The aim of such an approach is to focus on the evolution of a signal (e.g., heart rate) before, during, and after excitations taking the system out of its equilibrium (e.g., physical effort during cardiac stress testing). Dynamical modeling can be applied to a broad range of outcomes such as physiological processes in medicine and psychosocial processes in social sciences, and it allows to extract simple characteristics of the signal studied. The model we propose is based on a first order linear differential equation defined by three main parameters corresponding to the initial equilibrium value, the dynamic characteristic time, and the reaction to the excitation. In this paper, several estimation procedures for this model are considered and tested in a simulation study, that clarifies under which conditions accurate estimates are provided. Finally, applications of this model are illustrated using cardiology data recorded during effort tests.

# Dynamical system modeling of self-regulated systems undergoing multiple excitations: first order differential equation approach

## Introduction

Modern longitudinal studies have become more intensive, often producing 10 or more measurements per subject. These intensive longitudinal data (ILD, Walls & Schafer, 2006) stem from long cohort studies such as the Framingham cohort (Gordon, Kannel, Hjortland, & McNamara, 1978) or the Swiss HIV cohort (Elzi et al., 2010), but also from the use of electronic devices (e.g. online monitoring, personal digital assistant or cell phone reporting, Stone, Shiffman, Atienza, & Nebeling, 2007) or from electronic medical records. Following the emergence of these rich data, researchers have been able to examine changes over long time periods (Semba et al., 2005; West, Tan, Habermann, Sloan, & Shanafelt, 2009), for instance trajectories of disabilities (Gill, Gahbauer, Han, & Allore, 2015) and self-rated health (Cullati, Rousseaux, Gabadinho, Courvoisier, & Burton-Jeangros, 2014). Alternatively, some studies examined recurring cycles and patterns, such as the stability of the circadian temperature rhythm (Czeisler et al., 1999), the patterns of cigarette smoking events over several weeks (Shiffman & Rathbun, 2011), or the capacity of self-regulated systems to maintain stability through change and adapt themselves to environmental change (allostasis) (McEwen & Wingfield, 2003). An interesting type of evolution over time that can be assessed using ILD is homeostasis, that is the return to equilibrium of a given self-regulated system. Homeostasis is of particular interest in social science and medical research (Chrousos & Gold, 1992; Cullati, Kliegel, & Widmer, n.d.; Marieb & Hoehn, 2009; Toch & Hastorf, 1955) because it underlies many physiological processes, including regulation of body temperature, composition of the extracellular fluid (e.g., blood sugar level, carbon dioxide, oxygen) and gene regulation (Marieb & Hoehn, 2009). Homeostasis has also been postulated to underlie psychosocial regulation of stress (McEwen, 2006; Perry & Pollard, 1998), with individuals showing a tendency to stay at a certain level of stress, potentially generating artificial stress if the existing level of stress is insufficient (Bishop & Zito, 2013).

Since the 17th century and the work of Newton and Leibnitz (Newton, John Adams Library (Boston Public Library) BRL, Colson, & Adams, 1736), the concept of derivative (initially called fluxion by Newton) has been central to the development of modern science. A derivative is a quantification of a variable change in function of another variable. Derivation can be performed at different order: a second derivative is a measure of the variation of the first derivative, a third derivative is a measure of the variation of the second derivative, etc. In particular, the first derivative of a time dependent variable quantifies its instantaneous change over time (Courant & John, 1999), while the second derivative measures the instantaneous change over time of the first derivative. More precisely, considering a variable $Y(t)$ depending on the time $t$, the first derivative, written $Y'(t)$ in mathematics, $\dot{Y}(t)$ or $\frac{dY}{dt}(t)$ in physics or associated fields, is:

$$\dot{Y}(t) = \lim_{h \to 0} \frac{Y(t+h) - Y(t)}{h}$$

The first derivative of a continuous variable is the rate of the variable change between two points separated by a time $h$ when this time tends to 0. It corresponds to the tangent at point $t$ of the $Y(t)$ curve, and if $Y(t)$ measures the position in time of a system, then the first derivative $\dot{Y}(t)$ is its speed.

And the second derivative, written $Y''(t)$, $\ddot{Y}(t)$ or $\frac{d^2Y}{dt^2}(t)$ is the derivative of the first derivative, that is:

$$\ddot{Y}(t) = \lim_{h \to 0} \frac{\dot{Y}(t+h) - \dot{Y}(t)}{h}$$

The second derivative of a continuous variable is the rate of the first derivative change between two points separated by a time $h$ when this time tends to 0. It corresponds to the curvature at point $t$ of the $Y(t)$ curve, and if $Y(t)$ measures the position in time of a system, then the second derivative is its acceleration.

In 1998, the seminal work of Boker and Graham (Boker & Graham, 1998; Boker & Nesselroade, 2002) suggested the use of second order differential equations to study the link between an outcome, its first derivative and its second derivative.

This equation, usually used in physics to describe the behavior of an oscillator out of equilibrium, has been applied by the authors to explore the fluctuations of cigarettes and alcohol use in a cohort study of adolescent substance abuse. Various estimation procedures of the differential equation coefficients have been developed since then.

Among them, two-stage approaches (Chow, Bendezú, Cole, & Ram, 2016), consisting in first estimating the derivatives from the observed values with methods such as the initial Local Linear Approximation (LLA; Bisconti, Bergeman, & Boker, 2004; Boker & Nesselroade, 2002), its further development, the generalized linear local approximation (GLLA; Boker & Graham, 2010), the generalized orthogonal local derivative (GOLD; Deboeck, 2010), or a functional data analysis (FDA; Ramsay & Silverman, 2005; Trail et al., 2014) based approach, to then estimate their relation with the variable through a linear model. Several one-step procedures exist, such as the Latent differential equation modelling (LDE; Boker, Neale, & Rausch, 2004) approach based on structural equation modelling, the wide variety of estimation methods based on state space modeling stemming from econometrics using structural equation modeling techniques or applying the Kalman filter or further developments (Chow, 2018; Chow, Lu, Sherwood, & Zhu, 2016; Driver, Oud, & Voelkle, 2017; J. Oud, 2007) and the methods performing nonlinear regression directly from the analytic solutions of differential equations (ASDE, Hu & Treinen, 2019). As a consequence of this diversity of methods, numerous studies exist in the literature based on dynamical models, such as studies of the intraindividual variability (Deboeck, Montpetit, Bergeman, & Boker, 2009; Klipker, Wrzus, Rauers, Boker, & Riediger, 2017), first-order differential equations (Deboeck & Bergeman, 2013; Oravecz, Tuerlinckx, & Vandekerckhove, 2009; Oud, 2007), second order differential equations (Boker & Nesselroade, 2002; Chow et al., 2018; Chow, Lu, et al., 2016; Deboeck, Boker, & Bergeman, 2008; Klipker et al., 2017; Steele & Ferrer, 2011), coupled differential equations (Boker & Graham, 1998; Boker & Nesselroade, 2002; Butner, Amazeen, & Mulvey, 2005), and nonlinear differential equations (Butner et al., 2005; Chow, Bendezú, et al., 2016; Finan et al., 2010). However, most of these models ignore the potential presence excitations or inputs (shocks, adverse events, health hazards, etc.) altering the dynamics and pushing the observed outcome out of equilibrium. By excitation we designate any exogeneous process different from the variable studied or its derivative and appearing as a time-dependent covariate of the differential equation. This is known as the source term of the differential equation in physics (Harper, 1976) because it is the source of the movement of the system studied. Some assumptions can be made to ignore the excitations while performing a dynamical analysis: it can be supposed that the variable studied is not affected by any excitation during the measurement. In this case, the variable must be out of equilibrium at the beginning of the measurement (if not, there are no dynamics at all). Another approach is to consider that the excitations are brief compared to the time between two measurements and that few measurement points are affected. The effects of the excitation (for example phase shifts) can in this case be partially filtered by analysis processes (Bergeman & Boker, 2015). Finally, the excitation process can be considered as unknown and randomly distributed and can thus take the form of a dynamical noise integrated in the stochastic differential equation formalism (Oud & Delsing, 2010). But apart from these cases, ignoring this term in the differential equation can not only induce large errors in the estimation, but also deprive the researcher of a way of studying the link between the perturbation and the dynamics of the outcome.

Although this subject has been studied since decades in econometrics (Canova, Ciccarelli, & Ortega, 2012; Hamerle, Singer, & Nagl, 1993; Warsono, Russels, Wamiliana, Widiarti, & Usman, 2019), attempts in the behavioral field to account explicitly for the excitations perturbing the system in continuous time models are still recent (Deboeck & Bergeman, 2013; Driver et al., 2017; Timms, Rivera, Collins, & Piper, 2014; Trail et al., 2014).

For instance, the work of Trail and co-authors proposes several differential equation models to estimate the evolution of craving after smoking cessation. In this work, the action of stopping smoking is considered as an excitation mechanism included in the differential equation model. They use a two-step estimation procedure to estimate the parameters of the model, estimating first the derivatives of the variable through an FDA analysis and then using a mixed effect regression.

The lack of systematic simulation necessary to assess the conditions of use, the interpretation and the statistical characteristics of the methods including exogeneous inputs has been partially filled by a recent simulation study (Hu & Huang, 2018). However, the focus of this work on a particular case of differential equation (second order differential equation) and on a restricted case of exogenous mechanism (oscillatory excitation oscillating at a frequency equal to the natural frequency of the system) still leaves a gap that this article will try to fill.

The aim of this article is to extensively describe and characterize a two-step method similar to the one proposed by Trail and co-authors (Trail et al., 2014) to analyze the dynamics of return to equilibrium in self-regulated systems experiencing multiple excitations. This method will be compared to several other estimation procedures. We will focus on systems with a dynamic following a first order differential equation perturbed by a known excitation term. This simple differential equation is of high interest because it can apply to a wide range of systems and provide easily interpretable coefficients. Furthermore, a simple self-regulated system periodically set away from its equilibrium can also present oscillatory-like appearance that could be simply and accurately described by a first order differential equation considering multiple excitations.

This paper includes a description of the underlying mathematical model and estimation method chosen to fit its parameters and an extensive simulation study including the statistical characteristics of the estimated coefficients and a comparison of these results with other estimation procedures. The code used for the simulation study is provided as supplementary material, and the estimation method mentioned has been embedded in a dedicated R package on CRAN (Mongin, Uribe, & Courvoisier, 2019). Finally, we apply the method on a dataset of cardiological data recorded during effort tests, in order to illustrate the novelty that our approach could bring.

## A multiple shocks differential equation model

The general idea of a first order differential equation can be described as follows: given a variable $Y$ measuring a signal (e.g., affective state, biological measurement) dependent on time and with an equilibrium value $\mathrm{Y_{eq}}$, the first order differential equation with constant coefficients describes a linear relation between $Y$ and its first derivative $\dot{Y}$ (its instantaneous change over time; Courant & John, 1999):

$$\dot{Y}(t) + \gamma(Y(t) - Y_{eq}) = 0 \tag{1}$$

This differential equation, called homogeneous first order differential equation, is studied and used in various scientific fields since centuries (Feynman, Leighton, & Sands, 2011; Newton et al., 1736). For example, this equation describes the evolution of a population of radioactive atoms having a disintegrating rate of $\gamma$, the evolution of a bank account with pay taxes that is a percentage $\gamma$ of the account balance, or the evolution of a drug concentration in the body being eliminated at a rate of $\gamma$. Considering the last example, the differential equation simply states that the change in drug concentration over time (the derivative $\dot{Y}(t)$) is a fraction per amount of time ($\gamma$, lets say 1% every hour ) of the total drug concentration ($Y(t)$). The solution of this homogeneous equation is of the form:

$$Y(t) = (Y_0 - Y_{eq})e^{-\gamma t} + Y_{eq} = (Y_0 - Y_{eq})e^{-\frac{t}{\tau}} + Y_{eq} \tag{2}$$

where $Y_0 = \mathrm{Y(t = 0)}$ is the initial value of $Y$. The coefficient $\gamma$ is the decay rate and is linked to the decay time $\tau$:

$$\gamma = \frac{1}{\tau} \tag{3}$$

The system characteristic time or decay time, also known as mean lifetime when $Y$ designates a probability (Feynman et al., 2011), is the time constant of the exponential behavior that is the solution to equation (1). In equation 2, when replacing t = $\tau$:

$$Y(\tau) = \frac{Y_0 - Y_{eq}}{e} + Y_{eq} \simeq 37\%\left(Y_0 - Y_{eq}\right) + Y_{eq} = Y_0 - 63\%(Y_0 - Y_{eq})$$

In other words, $\tau$ is the time needed for $Y$ to reach 63% of its absolute change (see Figure 1, bottom right panel, where the solution of a first differential equation is represented for various $\tau$). As an example, a decay rate $\gamma$ of 1% per hour corresponds to a decay time of 100 hours. The homogeneous equation (1) describes the evolution of a signal already out of equilibrium and going back to equilibrium. But it does not consider the possible excitations that can occur to put the system out of equilibrium (in the example of drug concentration, the intake of a new dose of the drug). In this case, the equation becomes:

$$\dot{Y}(t) + \gamma(Y(t) - Y_{eq}) = \gamma K u(t) \tag{4}$$

where $u(t)$ is a time dependent excitation (i.e. exogenous input, shock), and $K$ is the gain (term coined from control engineering; Timms et al., 2014; Trail et al., 2014). $u(t)$, the excitation, is a vector of same length as $Y$ and could be for example a drug intake per amount of time (flow of drug) if $Y$ is a drug concentration.

The gain is the ratio between an increase of $u(t)$ and the increase that the variable will reach at steady-state (see Figure 1, top right, where the dynamic of a variable driven by the first order differential equation and experiencing a step excitation is represented for various gains). To clarify this point, let us consider that $Y$ starts at equilibrium and undergoes a constant excitation $U$. The steady state corresponds to the situation where the variable $Y$ does not change anymore with time, that is by definition $\dot{Y}(t) = 0$. The equation 4 reads in this case:

$$Y_{steady-state} = Y_{eq} + KU$$

$$K = \frac{Y_{steady-state} - Y_{eq}}{U}$$

The gain is thus the ratio between the magnitude of the system change and the excitation that caused it.

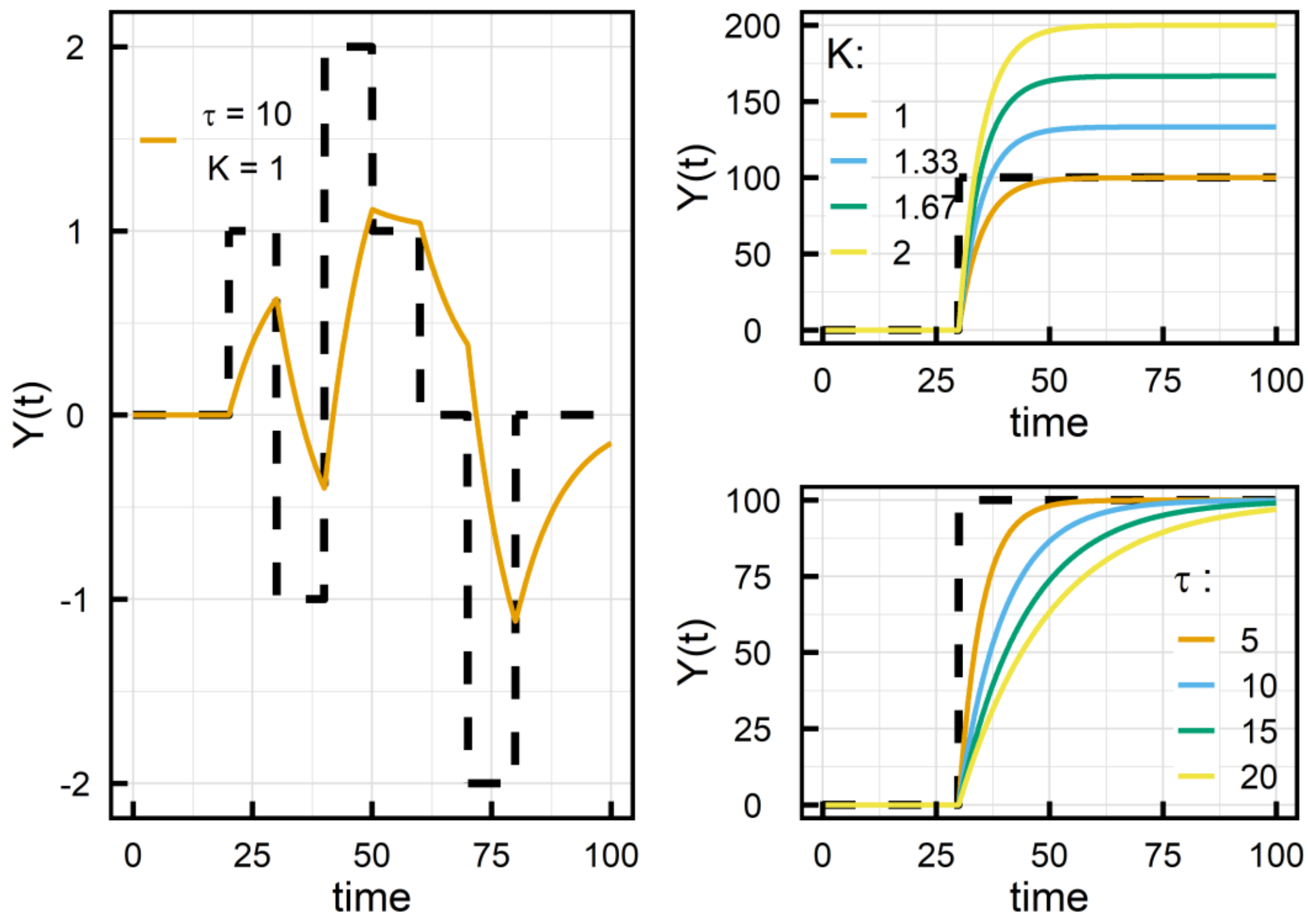

Figure 1: Examples of dynamics driven by a first order differential equation with known excitation (dashed lines). Left panel: decay time of $\tau = 10$, gain $K = 1$ and multiple excitations. Right top panel: Influence of the gain $K$ on the dynamics caused by a step excitation with a decay time $\tau = 5$. Right bottom panel: Influence of the decay time $\tau$ on the dynamics caused by a step excitation with a gain $K = 1$. Equilibrium value is 0 for all curves.

To present a simple case, let's consider that $u(t)$ is constant $(u)$ for any time $t > 0$. Considering that $Y(t = 0) = Y_{eq}$ for illustrative purposes, the solution of equation 4 has the form:

$$Y(t) = Ku\left(1 - e^{-\frac{t}{\tau}}\right) + Y_{eq} \tag{5}$$

The signal $Y(t)$ will in this case be an exponential increase (with decay time $\tau$) saturating at its steady state value $Ku + Y_{eq}$ (see Figure 1 bottom). Considering again our drug example, for a hypothetical injection of a quantity $u$ of drug per amount of time, the drug concentration in the body would reach a maximum value $Ku$ above its equilibrium (corresponding to the steady state where the body eliminates as much drug as what is injected) with a characteristic time $\tau$ (see Figure 1, right panel). Figure 2 represents the response to excitations of a system governed by a first order differential equation. The homogeneous differential equation (equation 1) is then valid only when the excitation term is null (see the white bands in Figure 2). Thus, it appears necessary to take into account the excitations occurring during the study in order to correctly estimate the decay time. Furthermore, including the excitation term in the equation allows to determine the gain $K$, thus quantifying the relationship between the source of the dynamics and its effect.

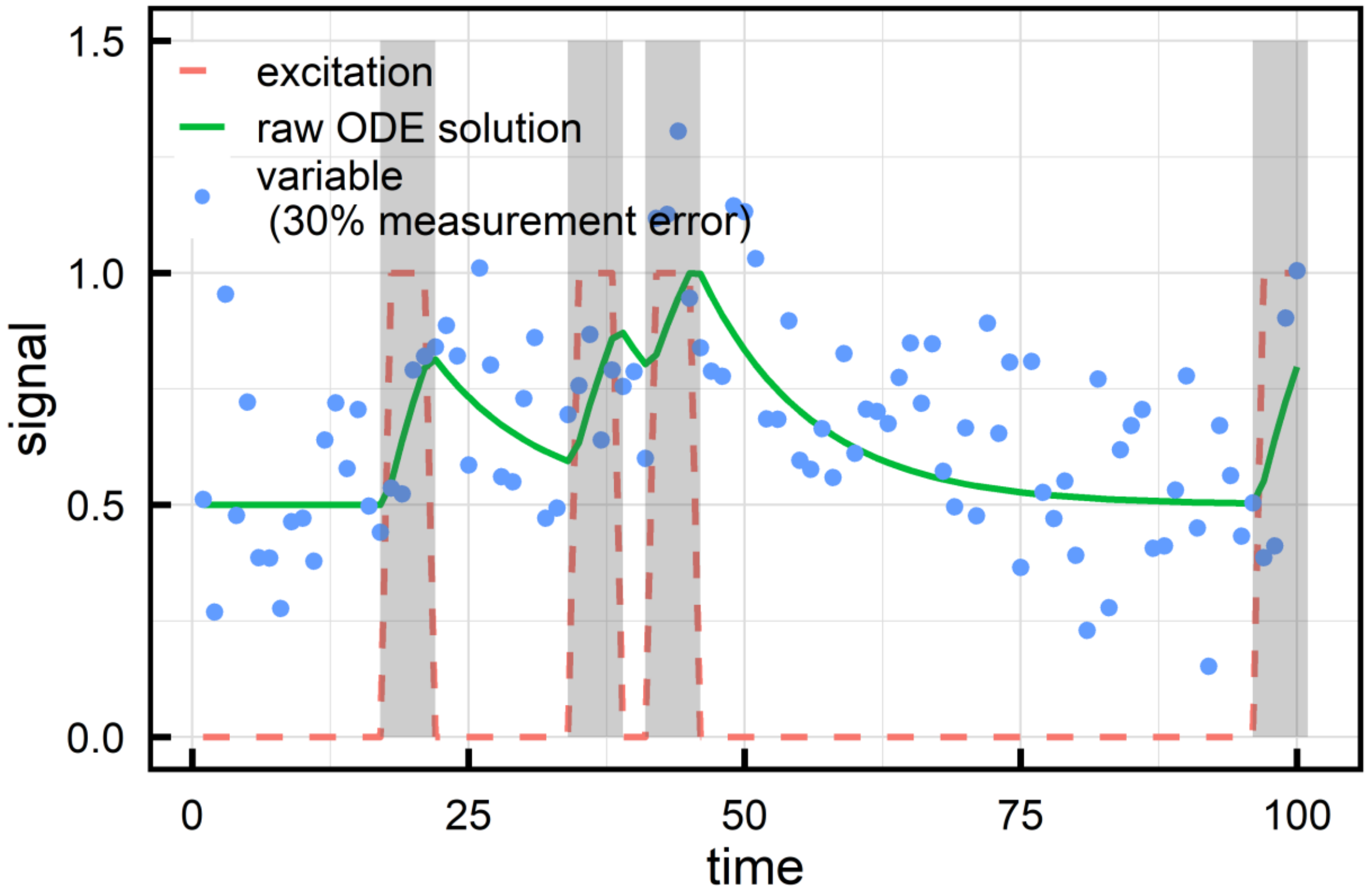

*Figure 2:* Variable (points) with 30% measurement error following the solution of a first order differential equation (solid line) provoked by a known excitation (dashed solid line). The equilibrium value is 0.5, the gain $K = 1$ and the decay time $\tau = 10$. The grey bands are the zones where the homogeneous differential equation (that is without considering excitation) is not valid.

## Estimation

A wide range of possible techniques exists to estimate the coefficients of a differential equation in dynamical modeling. In this paper, we will compare several methods to solve the differential equation model studied.

### Two step estimation procedures

Briefly, this method consists in first estimating the derivatives from the measurements and then performing a regression to estimate the linear relation between the derivatives and the variable. The estimation of the derivatives can be obtained with different methods (Chow, Bendezú, et al., 2016; Deboeck, 2010; Trail et al., 2014). We chose to compare the most popular: the Generalized Linear Local Approximation GLLA method (Boker & Graham, 2010; Chow, Bendezú, et al., 2016; Chow, Ferrer, & Hsieh, 2011) and the Functional Data Analysis (FDA) regression spline method (Chow, Bendezú, et al., 2016; Trail et al., 2014).

In brief, GLLA allows the calculation of derivatives using the information of a certain number of measurement points (called the embedding dimension d) spaced by $\tau_d$ occasions. We always considered the $d$ points to be adjacent (i.e. the spacing between successive lag of the embedding matrix is set to $\tau_d = 1$; Chow, Bendezú, et al., 2016). GLLA consists in multiplying the time-embedded data matrix $\mathbf{Y}$ by a matrix $\mathbf{W}$ containing the differencing weights derived from a Taylor expansion of $Y$. Boker et al. (Boker, Deboeck, Edler, & Peel, 2011) showed that $\mathbf{W}$ can be written as

$$\mathbf{W} = \mathbf{L}(\mathbf{L}'\mathbf{L})^{-1}$$

Where $\mathbf{L}$ is a $d \times (m+1)$ matrix, with $m$ the highest derivative order calculated, such that the jth column is given by:

$$L_j = \frac{[dt(\nu - \bar{\nu})]^{j-1}}{(j-1)!} \ , \ \nu = [1, \dots, d]$$

With $dt$ the time between successive points, $\nu$ a vector of integer from 1 to the embedding dimension $d$, and $\bar{\nu}$ its mean. The estimation of the derivative depends then on the embedding dimension chosen but also on the maximum derivative order considered when calculating $\mathbf{W}$. Recent studies show that using $m$ higher than the derivative order used can reduce the bias of the derivative estimation (Boker, Moulder, & Sjobeck, 2019; Chow, Bendezú, et al., 2016). As this is out of the scope of this paper, we will set $m = 1$. In this case, $\mathbf{W}$ is a $d \times 2$ matrix, and the GLLA method is strictly equivalent to the variant proposed by Deboeck, namely the Generalized Orthogonal Local Derivative (GOLD) method (Deboeck, 2010).

The FDA regression spline method consists on generating a B-spline function that fits the outcome to be studied and then estimating the derivative of that function. In order for the generated B-spline function to be differentiable, it needs to be smooth. This is achieved through a penalty function controlled by a smoothing parameter. The method is implemented in R through the *smooth.spline* function (Perperoglou, Sauerbrei, Abrahamowicz, & Schmid, 2019) which defaults allow to modify only the value of the smoothing parameter to adjust the fit, thus being very easy to implement. This method is applicable to observation with irregular time intervals. The choice of the smoothing parameter used during the first step of the estimation procedure is a complicated subject and can influence greatly its results (Deboeck et al., 2008; Hu, Boker, Neale, & Klump, 2014). The authors chose to perform an optimization procedure to select the proper embedding dimension d for the GLLA method and the appropriate smoothing parameter $spar$ for the FDA spline method. It consists in performing the analysis of the variable studied for various embedding dimensions or smoothing parameter (the embedding dimension or smoothing parameter is the same for all individuals), to then choose the one leading to the maximal value of the $R^2$ calculated with the curve estimated from the fixed effect coefficients and the variable.

The second step of the two-step method is the estimation of the differential equation coefficients. The authors chose linear mixed effects regression (Bisconti et al., 2004; Chow, Bendezú, et al., 2016; Deboeck et al., 2008; Gasimova et al., 2014; Trail et al., 2014) to account for inter-individual variations on the coefficients. The consideration of multiple excitations is done by including the known variable measuring them as a time-dependent predictor. The regression reads:

$$\dot{y}_{ij} \sim (b_1 + u_{1j})y_{ij} + (b_2 + u_{2j})u_{ij} + (b_0 + b_{0j}) + e_{ij} \tag{6}$$

with the index $j$ accounting for the different individuals, $i$ for the time, $e_{ij}$ for the error term. $\dot{y}$ is the derivative estimated on $d$ points, $y$ and $u$ are the signal and the excitation averaged on $d$ points.

Note that we include a random intercept $(b_0 + b_{0j})$, a random slope over the signal $(b_1 + u_{1j})$, and a random slope over the excitation term $(b_2 + u_{2j})$.

Reshaping equation 4 allows to identify the different parameters:

$$\dot{y}(t) = -\gamma y(t) + \gamma K u(t) + \gamma y_{eq}$$

The decay rate for each individual is then $\gamma_j = -(b_1 + u_{1j})$, and the corresponding decay time is then $\tau_j = \frac{1}{\gamma_j}$, the gain $K_j = \frac{(b_2 + u_{2j})}{\gamma_j}$ (see previous section and Figure 1) and the equilibrium value $y_{eq} = \frac{(b_0 + u_{0j})}{\gamma_j}$. Note that the parameters estimated by the mixed effect regression are $\gamma_j$, $\gamma_j K_j$ and $\gamma_j y_{eq}$.

The implementation of this second step was done in R using the *lmer* function from the *lme4* package (Bates, Mächler, Bolker, & Walker, 2015). With the above estimated parameters, the estimated signal can be reconstructed for each individual $j$ by performing a numerical integration of the differential equation with calculated coefficients and the known excitation term using the function *lsoda* from the package *deSolve* (Soetaert, Petzoldt, & Setzer, 2010). The estimated signal allows us to calculate the $R^2$ of the adjustment, providing a quality measurement of the overall two-step estimation.

**One-step estimation procedures**

A popular method to estimate the coefficients of a differential equation from discrete measurements is the use, within the general framework of the state-space modeling, of the Kalman filter (Kalman, 1960) or its continuous-time and nonlinear extensions.

The state space model associated with the first order differential equation presented can be written as:

$$\frac{dx(t)}{dt} = \gamma x(t) + \begin{pmatrix} \gamma K & \gamma Y_{eq} \end{pmatrix} \begin{pmatrix} u(t) \\ 1 \end{pmatrix} + d_t$$

where $x(t)$ is the latent variable undergoing the dynamics and $d_t$ a random Gaussian variable accounting for dynamical noise.

The measurement equation linking $x(t)$ with the measured variable $Y$ is the following:

$$Y(t) = x(t) + m_t$$

where $m_t$ is a random Gaussian variable accounting for the measurement error.

We considered in this article two slightly different estimations procedures based on tools recently released as R packages. The first one uses the continuous-discrete extended Kalman filter and is embedded in the tool developed by Chow and co-workers: the R package *dynr* (Ou, Hunter, & Chow, 2019). The second one is based on the continuous time part of the *mxExpectationStateSpace* function of the popular *OpenMX* package in R (Neale et al., 2016). It uses a hybrid Kalman filter to produce expectations, consisting in a Kalman-Bucy filter for the prediction step (Kalman & Bucy, 1961) and the classical Kalman filter (Kalman, 1960) for the update step.

Finally, the one-step method called Analytic Solutions of Differential Equations (ASDE; Hu & Treinen, 2019) was also included. A more versatile version of this method using numerical solutions instead of analytical solutions has been previously studied in pharmacokinetics (Tornøe, Agersø, Jonsson, Madsen, & Nielsen, 2004) and has been published as the R package *nlmeODE* (Tornoe, 2012). Shortly, the method consists in generating a solution of the differential equation using numerical integration (with the *lsoda* function of the *DeSolve* package) to then estimate the subsequent nonlinear model using mixed effect regression (based on the popular *nlme* package).

## Simulation

We generated multi-excitation driven damped exponential signals for $N_{indiv}$ individuals, each with $N_{obs}$ observations equally spaced in time by 1 (arbitrary) unit. By doing so, the value of the decay time in time unit is equivalent to the number of points per decay time. Our simulation signals were generated using adaptative numerical integration (using the *lsoda* function of the *deSolve* package). Measurement error (i.e. intraindividual noise) was added to the data as a normal distributed shift centered on zero and of standard deviation being a given percentage of the maximum signal amplitude. In order to account for inter-individual differences, an inter-individual noise of 20% was added, by setting the decay time, the gain and the equilibrium value of the individuals to be distributed along a normal distribution with a standard deviation of 20% of their mean.

In this simulation, we varied the damping rate $\gamma$, the excitation shape, the number of observation $N_{obs}$, the number of individual of the panel data $N_{indiv}$, the equilibrium value and the measurement error (STN). The gain was set to 1 for all simulation conditions, as varying the gain is similar to varying the measurement error (an increase of gain increases the system response without affecting its shape, see Figure 1 right panel). In all conditions, it was assumed that the outcome was at equilibrium at $t = 0$. However, all the analysis presented here can be conducted similarly if the outcome is not at equilibrium when starting the measurement. Finally, in order to highlight the importance of the excitation mechanism, the estimations were also performed without taking the excitation term into account (i.e. based on the homogeneous equation (1)). The simulation conditions are summarized in Table 1, and the shapes of excitation considered are given in Figure 3.

For each simulation condition, the simulated signal was generated 1000 times, and analyzed with various estimation methods: several two-steps and three single-step estimation methods, described below.

Concerning the two-step estimation procedures, we compared the use of the GLLA and the FDA spline method to calculate the derivatives. In the second step, in order to illustrate the advantage of the linear mixed effects approach, the regression was also performed with a standard linear model (using the *lm* function in R) and with a generalized estimating equation model (using the *geeglm* function from the *geepack* package in R; Højsgaard, Halekoh, & Yan, 2005). In the latter case, we supposed an exchangeable correlation structure. Indeed, the correlations between the derivative, the variable and the excitation are constant in time within each individual in the differential equation. During the optimization procedure used to choose the smoothing needed to calculate the derivative, the embedding dimension $d$ was varied with a step of 2 between its minimal value 2 and 25 the range encompassing half of the observations, and the smoothing parameter between 0 and 1 with steps of 0.2 (these ranges are given in the smooth.spline function manual as the standard range of values for the smoothing parameter). The chosen embedding dimension (for GLLA) or smoothing parameter (for the FDA method) is then kept constant for all estimation runs of the given simulation condition. The result of such procedure is illustrated in the beginning of the result section.

For one-step procedures, the algorithms used by *dynr, mxExpectationStateSpace* and *nlmeODE* require the starting values of the initial value, equilibrium value, decay rate and gain. The starting initial value was set as the mean of the initial values of all individuals. The starting equilibrium value was set as the mean of the variable over all times, and the starting values of the decay rate $\gamma$ and of the parameter associated with the gain $\gamma K$ were randomly sampled in the interval [0.1,1] and [0.1,3] respectively at each simulation run. In the state space equation model associated with *dynr* and *mxExpectationStateSpace*, the dynamical noise was set to zero.

| simulation condition | $\gamma$ | excitation | $N_{obs}$ | $N_{indiv}$ | $y_{eq}$ | STN | regression method | homogeneous |
|---|---|---|---|---|---|---|---|---|
| 1 | 1/15 | 2steps | 50 | 50 | 0.5 | 30 | lmer | false |
| 2 | **1/5** | 2steps | 50 | 50 | 0.5 | 30 | lmer | false |
| 3 | **1/10** | 2steps | 50 | 50 | 0.5 | 30 | lmer | false |
| 4 | **1/20** | 2steps | 50 | 50 | 0.5 | 30 | lmer | false |
| 5 | 1/15 | **1step** | 50 | 50 | 0.5 | 30 | lmer | false |
| 6 | 1/15 | **3pnt** | 50 | 50 | 0.5 | 30 | lmer | false |
| 7 | 1/15 | **5pnt** | 50 | 50 | 0.5 | 30 | lmer | false |
| 8 | 1/15 | **10pnt** | 50 | 50 | 0.5 | 30 | lmer | false |
| 9 | 1/15 | 2steps | **30** | 50 | 0.5 | 30 | lmer | false |
| 10 | 1/15 | 2steps | 50 | **20** | 0.5 | 30 | lmer | false |
| 11 | 1/15 | 2steps | 50 | **100** | 0.5 | 30 | lmer | false |
| 12 | 1/15 | 2steps | 50 | 50 | **0.0** | 30 | lmer | false |
| 13 | 1/15 | 2steps | 50 | 50 | 0.5 | **10** | lmer | false |
| 14 | 1/15 | 2steps | 50 | 50 | 0.5 | **50** | lmer | false |
| 15 | 1/15 | 2steps | 50 | 50 | 0.5 | 30 | **lm** | false |
| 16 | 1/15 | 2steps | 50 | 50 | 0.5 | 30 | **gee** | false |
| 17 | 1/15 | 2steps | 50 | 50 | 0.5 | 30 | lmer | **true** |

Table 1: summary of the simulation conditions considered for the simulation study. Simulation condition 1 is the reference condition. The parameters that varied compared to the reference are shaded in grey.

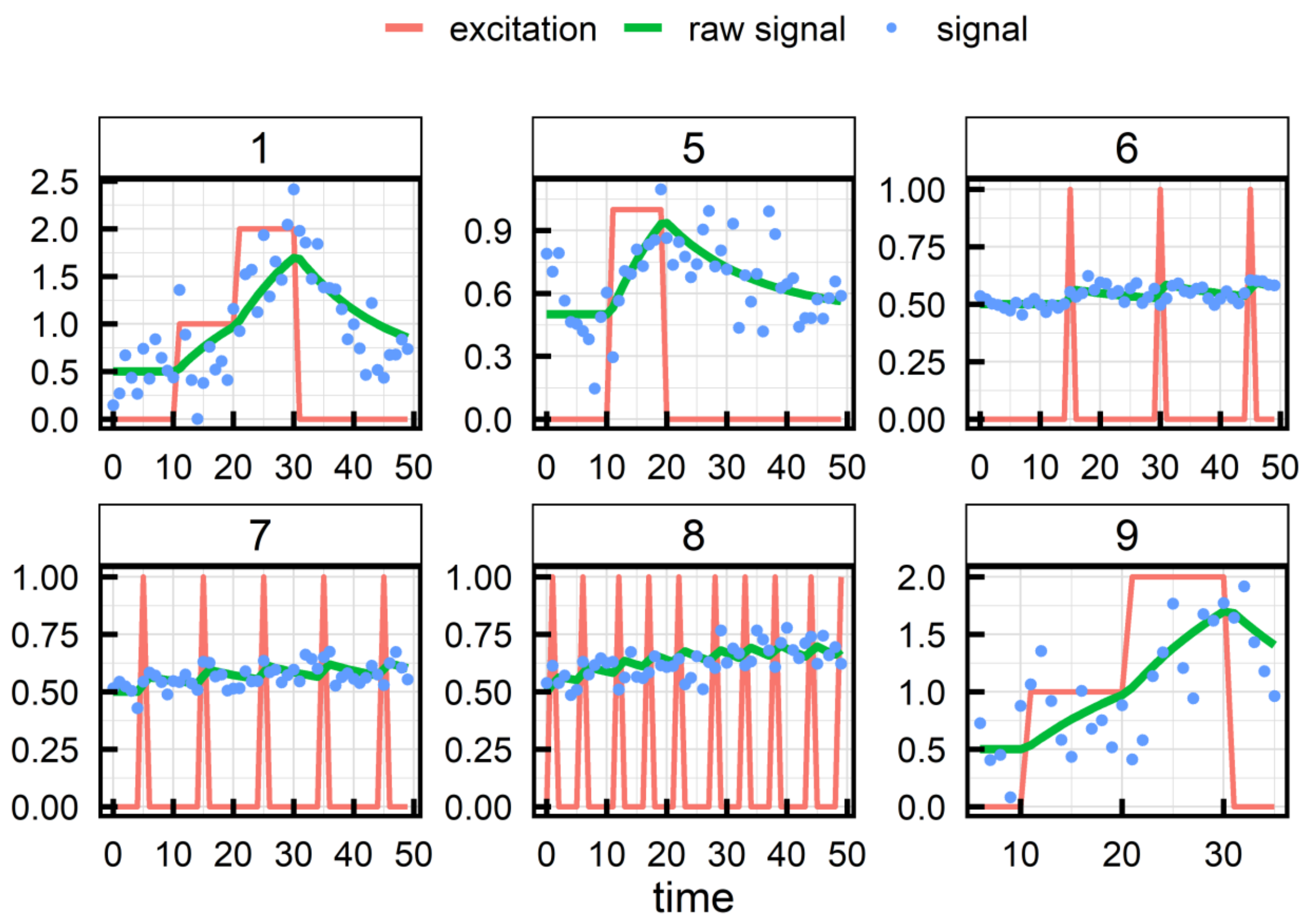

Figure 3: details of the excitation shape considered for the simulation conditions 1,5,6,7,8,9 (see Table 1).

**Statistical analysis**

For each estimation run, the median relative bias (with its [2.5 – 97.5%] range) of the three parameters estimated (the decay rate $\gamma$, $K\gamma$ and $y_{eq}\gamma$) by the different estimation methods considered was computed, as well as the percentage of estimations with an absolute relative bias under 10% (N10) and the coverage (cov). For each method and simulation condition, the median $R^2$ is calculated with the estimated outcome generated from the individual coefficients (R2r) or from the fixed coefficient (R2g) and the simulated signal without noise. The R code used to generate the simulation panels, estimate the coefficients and analyze the results is provided as supplementary material.

# Results

## Optimization procedure

To illustrate the merit of the procedure proposed to select the smoothing parameter needed for the estimation of the derivative within the two-step procedure, we present in figure 4 the mean $R^2$ obtained from the curve estimated with the fixed effect coefficients and the simulated signal with measurement noise, the mean estimated $K\gamma$, decay rate $\gamma$ and equilibrium parameter $y_{eq}\gamma$ obtained when analyzing 10 reference simulated panel with various embedding dimensions or smoothing parameter. It can be observed (top panel) that the $R^2$ increases with the smoothing used, reaches a temporary plateau and then decreases back, similarly to what Deboeck and co-authors (Deboeck et al., 2008) obtained using a slightly different two-step procedure. The vertical lines indicate the embedding dimension or smoothing parameter chosen, corresponding to the maximum $R^2$ for each method, while the horizontal dashed lines show the true initial values of the parameters. Figure 4 illustrates that the proposed “model free” optimization procedure yields the smoothing parameters leading to the less biased set of estimated parameters. It can be noted that the procedure proposed by Hu and co-authors (Hu et al., 2014), consisting in selecting the embedding dimension at which the estimate stabilizes, cannot be applied in our case because the stabilization does not occur.

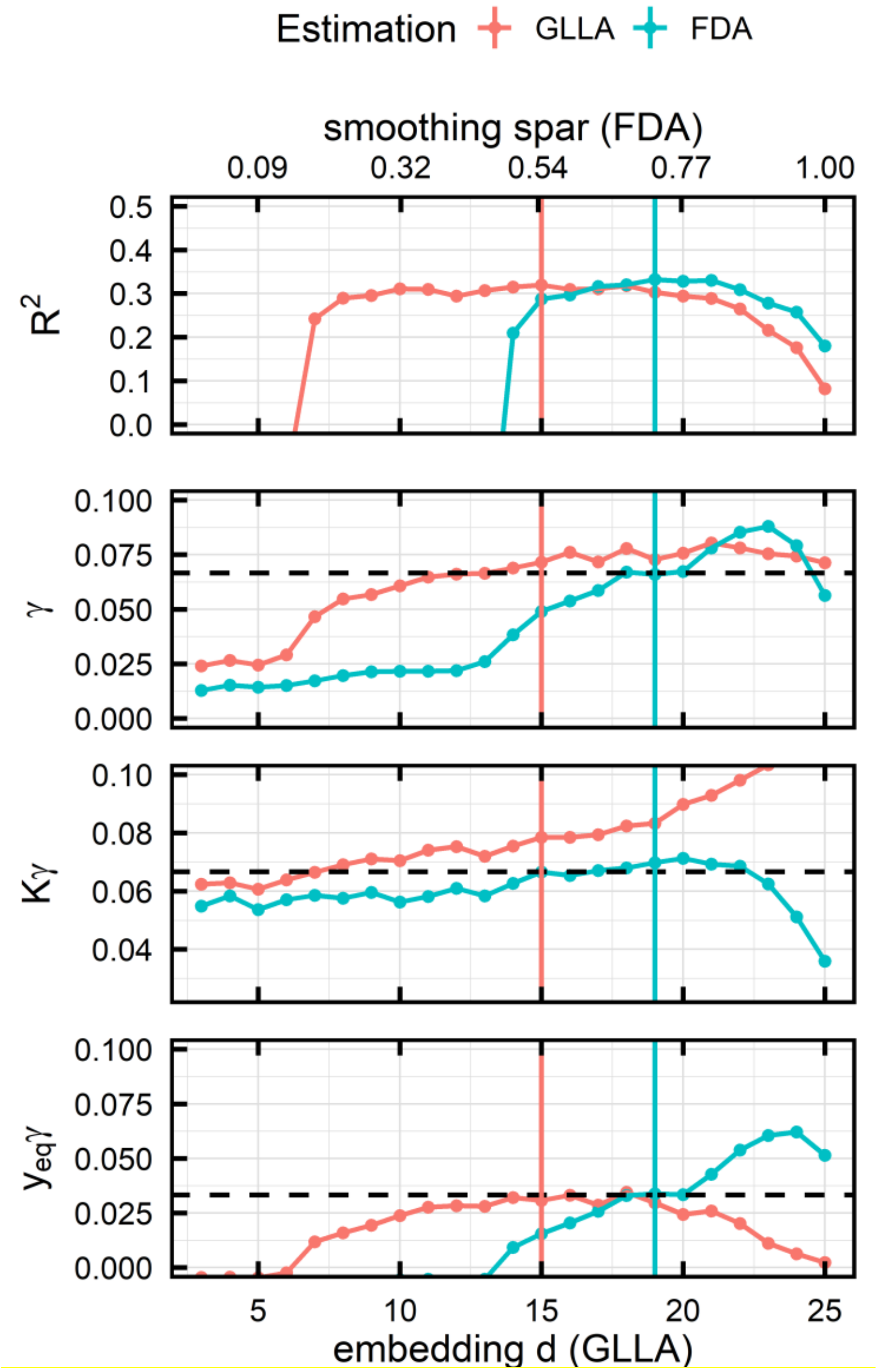

Figure 4: mean $R^2$ calculated between the curve estimated with the fixed effect coefficient and the variable studied (including measurement noise), the mean estimated gain K, decay time $\tau$ and equilibrium value $y_{eq}$ obtained when analyzing ten reference simulated panel for a range of embedding dimensions d or smoothing parameter spar. Dashed horizontal lines designate the true value of the parameters. Vertical lines indicate the embedding dimension or smoothing parameter corresponding to the maximum value of $R^2$

## Simulation results

In this section, we present the results regarding the estimation of the three parameters of the first order differential equation: the decay time of our signal, the gain and the equilibrium value, for each simulation condition and compare these results to the reference simulation condition. The complete set of simulation results can be found in the supplementary Table S1. The light grey cells correspond to the simulation condition where the estimated parameter was at least 50% of the time below 10% of absolute relative bias. The darker grey cells in the $R^2$ column correspond to $R^2$ above 0.8.

*Reference parameters*

| params | | $\gamma$ | | | $K\gamma$ | | | $y_{eq}\gamma$ | | | R2 | | d |
|---|---|---|---|---|---|---|---|---|---|---|---|---|---|
| | **Method** | **bias** | **N10** | **cov** | **bias** | **N10** | **cov** | **bias** | **N10** | **cov** | **R2r** | **R2g** | **d** |
| 1: Reference condition | DYNR | 2 [-12;2672] | 35.1 | 44.1 | 10 [-4;843] | 49.5 | 0.1 | 12 [-33;3796] | 13.9 | 40.9 | 0.37 | 0.37 | |
| | OPENMX | -7 [-15;0] | 66.4 | 84.1 | 0 [-5;8] | 93.6 | 0 | -16 [-38;5] | 28.5 | 79.5 | 0.45 | 0.45 | |
| | NLME | 2194 [383;3846] | 0.2 | 0.9 | 160 [-26;516] | 0.4 | 0 | 1749 [384;2747] | 0.4 | 1.4 | 0.50 | 0.10 | |
| | GLLA | 8 [-1;17] | 63.6 | 81.1 | 17 [9;25] | 14.0 | 0 | -7 [-31;17] | 37.1 | 91.8 | 0.79 | 0.45 | 15 |
| | FDA | -1 [-8;7] | 91.9 | 93.5 | 5 [-2;12] | 81.7 | 0 | -8 [-28;13] | 41.2 | 91.8 | 0.81 | 0.45 | 0.73 |
| 15: Homogeneous equat | DYNR | 43 [-250;57] | 0.1 | 0.1 | | | | 79 [50;97] | 0.0 | 0 | | | |
| | OPENMX | 8 [2;13] | 71.0 | 98.5 | | | | 179 [160;202] | 0.0 | 0 | | | |
| | GLLA | -78 [-82;-72] | 0.0 | 0 | | | | -11 [-24;5] | 43.6 | 70.9 | | | 13 |
| | FDA | -81 [-86;-76] | 0.0 | 0 | | | | -38 [-49;-24] | 0.6 | 2.5 | | | 0.82 |

Table 2: relative bias (bias), percentage of estimations with an absolute relative bias below 10% (N10) and coverage (cov) of the estimated coefficients (decay rate $\gamma$, parameters associated with the gain $K\gamma$ and with the equilibrium value $y_{eq}\gamma$) for the different estimation methods applied to the reference simulation condition (condition 1) and to the simulation condition where the excitation process is not taken into account during the analysis (condition 15). The median $R^2$ calculated with the estimated curve from the individual coefficients (R2r) and from the fixed effect coefficient (R2g) and the simulation signal without noise, as well as the embedding number or smoothing parameter used (d) are displayed. For condition 17, the gain is not estimated, and no estimated curve is generated, leading to empty cells for the $R^2$ and gain statistics.

Table 2 displays the results obtained with the different estimation methods for the reference simulation situation (simulation condition 1) and compares it with the same analysis when not taking the excitation process into account. *nlmeODE*, yields strongly overestimated parameters, still explaining 50% of the initial variance of the simulated panel data. *mxExpectationStateSpace* from *OpenMx* provides good estimates of the decay rate and parameter associated with gain $K\gamma$ and a slightly underestimated parameter $y_{eq}\gamma$ with an underestimated standard error. *Dynr* provides unbiased median estimates for the three parameters, but the bias distribution is strongly skewed towards high positive values. The two-step procedures estimate coefficients that explain around 80% of the initial variance. Looking at the relative bias of the estimated parameters, the FDA method yields unbiased results for the three parameters, and acceptable standard errors for decay rate $\gamma$ and parameter associated with equilibrium $y_{eq}\gamma$ leading to coverage between 80% and 100%. GLLA performs slightly worst by yielding an overestimated parameter associated with gain $K\gamma$.

When not taking into account the excitation process (simulation condition 17), the model misspecification leads to strong bias for the two-step methods and *dynr*, highlighting the importance of considering the potential exogeneous inputs when performing such analyses. Surprisingly, *mxExpectationStateSpace* is still able to provide a good estimate of the decay rate, although the equilibrium value is strongly biased.

*Effect of decay rate (or number of points per decay time)*

| params | | $\gamma$ | | | $K\gamma$ | | | $y_{eq}\gamma$ | | | R2 | | d |
|---|---|---|---|---|---|---|---|---|---|---|---|---|---|
| | Method | bias | N10 | cov | bias | N10 | cov | bias | N10 | cov | R2r | R2g | d |
| 2: $\gamma$ = 1/5 | DYNR | -8 [-16;720] | 20.6 | 16.1 | -4 [-13;437] | 40.9 | 27.9 | -5 [-29;996] | 17.6 | 30.6 | 0.61 | 0.61 | |
| | OPENMX | -12 [-18;-6] | 34.4 | 25.6 | -8 [-14;-2] | 65.6 | 0 | -17 [-33;2] | 27.2 | 47.8 | 0.66 | 0.66 | |
| | GLLA | -3 [-8;4] | 94.1 | 87.3 | 4 [-2;12] | 80.9 | 0 | -12 [-29;7] | 39.6 | 83.3 | 0.96 | 0.65 | 11 |
| | FDA | -10 [-16;-4] | 46.3 | 38 | -5 [-11;3] | 84.6 | 0 | -17 [-33;0] | 28.8 | 71.9 | 0.94 | 0.65 | 0.73 |
| 3: $\gamma$ = 1/10 | DYNR | -2 [-13;1728] | 36.9 | 38.1 | 4 [-6;736] | 53.0 | 0.8 | 1 [-29;2387] | 15.3 | 39.7 | 0.48 | 0.48 | |
| | OPENMX | -8 [-14;-1] | 63.7 | 64.4 | -2 [-8;5] | 92.2 | 0 | -16 [-33;6] | 26.5 | 66.6 | 0.53 | 0.53 | |
| | GLLA | -2 [-9;6] | 92.0 | 90.4 | 9 [1;16] | 59.6 | 0 | -14 [-36;6] | 31.5 | 77.7 | 0.93 | 0.54 | 11 |
| | FDA | -4 [-10;3] | 90.1 | 86.4 | 3 [-3;10] | 89.7 | 0 | -12 [-31;9] | 35.8 | 83.9 | 0.89 | 0.54 | 0.73 |
| 1: Reference condition: $\gamma$ = 1/15 | DYNR | 2 [-12;2672] | 35.1 | 44.1 | 10 [-4;843] | 49.5 | 0.1 | 12 [-33;3796] | 13.9 | 40.9 | 0.37 | 0.37 | |
| | OPENMX | -7 [-15;0] | 66.4 | 84.1 | 0 [-5;8] | 93.6 | 0 | -16 [-38;5] | 28.5 | 79.5 | 0.45 | 0.45 | |
| | NLME | 2194 [383;3846] | 0.2 | 0.9 | 160 [-26;516] | 0.4 | 0 | 1749 [384;2747] | 0.4 | 1.4 | 0.50 | 0.10 | |
| | GLLA | 8 [-1;17] | 63.6 | 81.1 | 17 [9;25] | 14.0 | 0 | -7 [-31;17] | 37.1 | 91.8 | 0.79 | 0.45 | 15 |
| | FDA | -1 [-8;7] | 91.9 | 93.5 | 5 [-2;12] | 81.7 | 0 | -8 [-28;13] | 41.2 | 91.8 | 0.81 | 0.45 | 0.73 |
| 4: $\gamma$ = 1/20 | DYNR | 3108 [-14;3649] | 25.0 | 40.4 | 484 [-2;911] | 40.9 | 0 | 4105 [-32;5200] | 10.3 | 39.9 | 0.01 | 0.01 | |
| | OPENMX | -8 [-16;1] | 60.8 | 92 | 2 [-5;9] | 91.6 | 0 | -19 [-40;6] | 24.0 | 86.8 | 0.38 | 0.38 | |
| | GLLA | 5 [-4;15] | 72.8 | 88.3 | 17 [9;25] | 12.9 | 0 | -13 [-38;14] | 29.9 | 86.7 | 0.46 | 0.38 | 15 |
| | FDA | 0 [-9;9] | 87.5 | 93.9 | 6 [-2;12] | 78.4 | 0 | -6 [-29;19] | 37.1 | 91.8 | 0.63 | 0.39 | 0.73 |

Table 3: relative bias (bias), percentage of estimations with an absolute relative bias below 10% (N10) and coverage (cov) of the estimated (decay rate $\gamma$, parameters associated with the gain $K\gamma$ and with the equilibrium value $y_{eq}\gamma$) for the different estimation methods and for different simulated decay time (2: $\tau = 5$, 3: $\tau = 10$, 1: $\tau = 15$, 4: $\tau = 20$). The median $R^2$ calculated with the estimated curve from the individual coefficients (R2r) and from the fixed effect coefficient (R2g) and the simulation signal without noise, as well as the embedding number or smoothing parameter used (d) are displayed.

*dynr*, yield strongly biased estimates for short decay rates ($\gamma = 1/20$), and seems to improve its estimation for higher decay rates.

Varying the decay rate from 1/5 to 1/20 (see Table 3) shows that the two-step procedures and the estimation performed by *mxExpectationStateSpace* one need at least 5 points per decay time to provide accurate estimate of decay rate. Indeed, for this condition, the FDA and *mxExpectationStateSpace* methods starts to yield underestimated decay rates. GLLA seems to produce proper estimates for decay rate as high as 1/5.

*Effect of equilibrium value*

| | params | $\gamma$ | | | $K\gamma$ | | | $y_{eq}\gamma$ | | | R2 | | d |
|---|---|---|---|---|---|---|---|---|---|---|---|---|---|
| | Method | bias | N10 | cov | bias | N10 | cov | bias | N10 | cov | R2r | R2g | d |
| 1: Reference condition | DYNR | 2 [-12;2672] | 35.1 | 44.1 | 10 [-4;843] | 49.5 | 0.1 | 12 [-33;3796] | 13.9 | 40.9 | 0.37 | 0.37 | |
| | OPENMX | -7 [-15;0] | 66.4 | 84.1 | 0 [-5;8] | 93.6 | 0 | -16 [-38;5] | 28.5 | 79.5 | 0.45 | 0.45 | |
| | NLME | 2194 [383;3846] | 0.2 | 0.9 | 160 [-26;516] | 0.4 | 0 | 1749 [384;2747] | 0.4 | 1.4 | 0.50 | 0.10 | |
| | GLLA | 8 [-1;17] | 63.6 | 81.1 | 17 [9;25] | 14.0 | 0 | -7 [-31;17] | 37.1 | 91.8 | 0.79 | 0.45 | 15 |
| | FDA | -1 [-8;7] | 91.9 | 93.5 | 5 [-2;12] | 81.7 | 0 | -8 [-28;13] | 41.2 | 91.8 | 0.81 | 0.45 | 0.73 |
| 12: $Y_{eq} = 0$ | DYNR | 2281 [-13;2676] | 27.0 | 35.9 | 472 [-3;851] | 44.1 | 0.1 | 0.00 [-0.00;0.00] | 99.5 | 36.1 | 0.07 | 0.07 | |
| | OPENMX | -2 [-10;7] | 82.5 | 93.7 | -2 [-9;5] | 89.0 | 0 | 0.00 [0.00;0.00] | 100.0 | 97.2 | 0.45 | 0.45 | |
| | GLLA | 2 [-6;11] | 83.2 | 92.6 | 14 [6;22] | 27.3 | 0 | -0.00 [-0.00;0.00] | 100.0 | 76.9 | 0.84 | 0.45 | 13 |
| | FDA | -1 [-8;7] | 91.2 | 93.8 | 5 [-2;12] | 81.1 | 0 | -0.00 [-0.00;0.00] | 100.0 | 89.7 | 0.81 | 0.45 | 0.73 |

Table 4: relative bias (bias), percentage of estimations with an absolute relative bias below 10% (N10) and coverage (cov) of the estimated coefficients (decay rate $\gamma$, parameters associated with the gain $K\gamma$ and with the equilibrium value $y_{eq}\gamma$) for the different estimation methods and for two equilibrium values (1:$y_{eq} = 0.5$ ,12 : $y_{eq} = 0$. In the latter case, the bias indicated is the equilibrium value). The median $R^2$ calculated with the estimated curve from the individual coefficients (R2r) and from the fixed effect coefficient (R2g) and the simulation signal without noise, as well as the embedding number or smoothing parameter used (d) are displayed.

When simulating panel data with a null equilibrium value, *dynr* seems to produce more high overestimated value of the parameter, leading to an important median bias value, while all other methods provide good estimate of the equilibrium associated parameter $y_{eq}\gamma$ and perform similarly to the reference condition for the two others estimated parameters.

*Effect of the excitation shape and number*

| | params | $\gamma$ | | | $K\gamma$ | | | $y_{eq}\gamma$ | | | R2 | | d |
|---|---|---|---|---|---|---|---|---|---|---|---|---|---|
| | Method | bias | N10 | cov | bias | N10 | cov | bias | N10 | cov | R2r | R2g | d |
| 1: Reference condition | DYNR | 2 [-12;2672] | 35.1 | 44.1 | 10 [-4;843] | 49.5 | 0.1 | 12 [-33;3796] | 13.9 | 40.9 | 0.37 | 0.37 | |
| | OPENMX | -7 [-15;0] | 66.4 | 84.1 | 0 [-5;8] | 93.6 | 0 | -16 [-38;5] | 28.5 | 79.5 | 0.45 | 0.45 | |
| | NLME | 2194 [383;3846] | 0.2 | 0.9 | 160 [-26;516] | 0.4 | 0 | 1749 [384;2747] | 0.4 | 1.4 | 0.50 | 0.10 | |
| | GLLA | 8 [-1;17] | 63.6 | 81.1 | 17 [9;25] | 14.0 | 0 | -7 [-31;17] | 37.1 | 91.8 | 0.79 | 0.45 | 15 |
| | FDA | -1 [-8;7] | 91.9 | 93.5 | 5 [-2;12] | 81.7 | 0 | -8 [-28;13] | 41.2 | 91.8 | 0.81 | 0.45 | 0.73 |
| 5: Excitation: one step | DYNR | 2580 [-31;2804] | 1.7 | 10.1 | 392 [-3;504] | 25.8 | 0.1 | 2965 [-44;3329] | 1.2 | 9.5 | 0.02 | 0.02 | |
| | OPENMX | -14 [-22;-5] | 28.3 | 88.4 | -0 [-5;6] | 97.0 | 0 | -20 [-32;-6] | 17.4 | 85.3 | 0.30 | 0.30 | |
| | GLLA | -4 [-14;7] | 70.6 | 87.5 | 14 [8;23] | 20.8 | 0 | -12 [-27;5] | 39.3 | 78.7 | 0.79 | 0.29 | 11 |
| | FDA | -11 [-20;-1] | 40.7 | 58.1 | 2 [-4;10] | 90.3 | 0 | -16 [-30;-1] | 28.3 | 65.7 | 0.50 | 0.30 | 0.64 |
| 6: Excitation: 3 points | DYNR | 2650 [-176;2855] | 3.8 | 1.6 | 37 [-29;84] | 9.3 | 0 | 2683 [-193;3095] | 3.9 | 1.6 | 0.00 | 0.00 | |
| | OPENMX | -9 [-17;-0] | 58.4 | 99.8 | -14 [-20;-8] | 20.6 | 0 | -8 [-18;3] | 58.2 | 99.8 | 0.02 | 0.02 | |
| | GLLA | -11 [-25;1] | 43.5 | 70.8 | -73 [-78;-70] | 0.0 | 0 | -2 [-19;14] | 55.8 | 86.8 | 0.80 | 0.00 | 25 |
| | FDA | -26 [-36;-16] | 2.4 | 7.2 | -3 [-12;5] | 79.9 | 0 | -26 [-38;-15] | 4.8 | 18.1 | 0.81 | 0.02 | 0.73 |
| 7: Excitation: 5 points | DYNR | 2558 [-128;2694] | 4.4 | 4.1 | -47 [-68;45] | 5.4 | 0 | 2764 [-142;3124] | 3.1 | 4.1 | 0.00 | 0.00 | |
| | OPENMX | 4 [-7;15] | 72.3 | 100 | -24 [-31;-17] | 0.6 | 0 | 9 [-5;24] | 48.9 | 99.9 | 0.02 | 0.02 | |
| | GLLA | -17 [-51;-3] | 26.0 | 53.1 | -96 [-99;-93] | 0.0 | 0 | -1 [-43;18] | 47.3 | 82.8 | 0.28 | 0.00 | 25 |
| | FDA | -37 [-97;-18] | 3.4 | 15.3 | -13 [-20;-4] | 33.7 | 0 | -40 [-107;-15] | 5.1 | 25 | 0.00 | 0.01 | 0.55 |
| 8: Excitation: 10 points | DYNR | 2519 [-236;2646] | 1.1 | 2.3 | -75 [-100;18] | 7.1 | 0 | 3062 [-306;3380] | 1.4 | 2.2 | 0.00 | 0.00 | |
| | OPENMX | 11 [-1;23] | 46.0 | 99.7 | -44 [-53;-35] | 0.0 | 0 | 30 [13;47] | 6.9 | 99.5 | 0.03 | 0.03 | |
| | GLLA | -28 [-40;-15] | 4.5 | 10.7 | -73 [-77;-70] | 0.0 | 0 | -9 [-25;9] | 46.1 | 79.2 | 0.91 | 0.00 | 15 |
| | FDA | -8 [-21;4] | 53.8 | 79.5 | -99 [-112;-86] | 0.0 | 0 | 26 [8;47] | 11.5 | 53.2 | 0.00 | 0.00 | 0.64 |

Table 5: relative bias (bias), percentage of estimations with an absolute relative bias below 10% (N10) and coverage (cov) of the estimated coefficients (decay rate $\gamma$, parameters associated with the gain $K\gamma$ and with the equilibrium value $y_{eq}\gamma$) for the different estimation methods and for different excitation shapes (1: two steps, 5: one step, 6: 3 punctual excitations, 7: 5 punctual excitations, 8: 10 punctual excitations). The median $R^2$ calculated with the estimated curve from the individual coefficients (R2r) and from the fixed effect coefficient (R2g) and the simulation signal without noise, as well as the embedding number or smoothing parameter used (d) are displayed.

The effect of excitation shape and number is summarized in Table 5. When confronted to shorter excitations or punctual excitations, *dynr* yields strongly overestimated decay rates and parameter $K\gamma$.

The presence of a shorter unique step excitation (simulation condition 5) leads to an increased positive bias of the gain for the two-step procedures compared to the reference situation. GLLA is the procedure that performs best in this condition. When the outcome contains quick changes caused by the punctual excitations (simulation condition 6, 7 and 8, see Figure 3), the results are significantly different. In this case, *mxExpectationStateSpace* still delivers correct estimates of the decay rate, while two-step procedures underestimate the decay rate and the parameter associated with gain $K\gamma$. The $R^2$ associated with the estimation indicates that GLLA performs better than the spline method for shorter excitations. Increasing the number of excitations leads to an improvement of the estimation produced by GLLA.

*Effect of number of observations and sample size*

| params | | $\gamma$ | | | $K\gamma$ | | | $y_{eq}\gamma$ | | | R2 | | d |
|---|---|---|---|---|---|---|---|---|---|---|---|---|---|
| | **Method** | **bias** | **N10** | **cov** | **bias** | **N10** | **cov** | **bias** | **N10** | **cov** | **R2r** | **R2g** | **d** |
| 1: Reference condition | DYNR | 2 [-12;2672] | 35.1 | 44.1 | 10 [-4;843] | 49.5 | 0.1 | 12 [-33;3796] | 13.9 | 40.9 | 0.37 | 0.37 | |
| | OPENMX | -7 [-15;0] | 66.4 | 84.1 | 0 [-5;8] | 93.6 | 0 | -16 [-38;5] | 28.5 | 79.5 | 0.45 | 0.45 | |
| | NLME | 2194 [383;3846] | 0.2 | 0.9 | 160 [-26;516] | 0.4 | 0 | 1749 [384;2747] | 0.4 | 1.4 | 0.50 | 0.10 | |
| | GLLA | 8 [-1;17] | 63.6 | 81.1 | 17 [9;25] | 14.0 | 0 | -7 [-31;17] | 37.1 | 91.8 | 0.79 | 0.45 | 15 |
| | FDA | -1 [-8;7] | 91.9 | 93.5 | 5 [-2;12] | 81.7 | 0 | -8 [-28;13] | 41.2 | 91.8 | 0.81 | 0.45 | 0.73 |
| 9: 30 Observations | DYNR | 2260 [-112;2338] | 6.1 | 33.9 | 851 [-47;960] | 12.7 | 0 | 1693 [-96;2313] | 6.3 | 33.7 | 0.42 | 0.42 | |
| | OPENMX | -1 [-45;49] | 21.7 | 98.2 | -0 [-19;23] | 44.6 | 0 | -7 [-45;39] | 24.1 | 97.8 | 0.45 | 0.45 | |
| | GLLA | -106 [-116;-95] | 0.0 | 0 | -34 [-43;-25] | 0.0 | 0 | -101 [-116;-85] | 0.0 | 0 | 0.00 | 0.00 | 23 |
| | FDA | -99 [-118;-80] | 0.0 | 0.2 | -43 [-57;-28] | 0.3 | 0 | -87 [-118;-55] | 0.0 | 70.3 | 0.00 | 0.00 | 0.18 |
| 10: 20 individuals | DYNR | 7 [-16;2752] | 28.4 | 49 | 14 [-6;921] | 41.7 | 3.3 | 23 [-38;4002] | 12.9 | 46.9 | 0.34 | 0.34 | |
| | OPENMX | -7 [-18;5] | 57.4 | 92.7 | 0 [-9;13] | 76.7 | 0.1 | -16 [-47;16] | 23.6 | 88 | 0.46 | 0.46 | |
| | GLLA | -4 [-17;10] | 62.4 | 88.1 | 10 [-2;22] | 49.4 | 0 | -21 [-56;16] | 19.8 | 80.4 | 0.89 | 0.45 | 11 |
| | FDA | -2 [-12;11] | 69.7 | 92.3 | 5 [-6;16] | 69.8 | 0 | -8 [-43;28] | 27.3 | 89.6 | 0.84 | 0.46 | 0.73 |
| 11: 100 individuals | DYNR | 2282 [-11;2658] | 32.9 | 32.7 | 473 [-2;828] | 47.2 | 0 | 3047 [-28;3766] | 9.5 | 30.5 | 0.07 | 0.07 | |
| | OPENMX | -7 [-13;-1] | 72.1 | 71.5 | 1 [-4;6] | 96.9 | 0 | -16 [-30;-1] | 24.0 | 70 | 0.45 | 0.45 | |
| | GLLA | 7 [1;14] | 68.8 | 64.5 | 17 [11;23] | 5.2 | 0 | -7 [-24;10] | 50.1 | 88.8 | 0.74 | 0.45 | 15 |
| | FDA | -1 [-6;5] | 97.4 | 92 | 5 [0;10] | 88.8 | 0 | -7 [-22;8] | 53.7 | 88.7 | 0.79 | 0.45 | 0.73 |

Table 6: relative bias (bias), percentage of estimations with an absolute relative bias below 10% (N10) and coverage (cov) of the estimated coefficients (decay rate $\gamma$, parameters associated with the gain $K\gamma$ and with the equilibrium value $y_{eq}\gamma$) for the different estimation methods and for different individual and observation numbers (1: $N_{obs} = 50$, $N_{indiv} = 50$, 9: $N_{obs} = 30$, $N_{indiv} = 50$, 10: $N_{obs} = 50$, $N_{indiv} = 20$, 11: $N_{obs} = 50$, $N_{indiv} = 100$). ). The median $R^2$ calculated with the estimated curve from the individual coefficients (R2r) and from the fixed effect coefficient (R2g) and the simulation signal without noise, as well as the embedding number or smoothing parameter used (d) are displayed.

The decrease to 30 observations per individual (simulation condition 9) causes an increase of the absolute relative bias for all methods (see Table 6). On the contrary, the presence of only 20 individuals (simulation condition 10) still allows the two steps methods and *mxExpectationStateSpace* to deliver accurate results.

*Effect of measurement error*

| | params | $\gamma$ | | | $K\gamma$ | | | $y_{eq}\gamma$ | | | R2 | | d |
|---|---|---|---|---|---|---|---|---|---|---|---|---|---|
| | Method | bias | N10 | cov | bias | N10 | cov | bias | N10 | cov | R2r | R2g | d |
| 13: STN = 10% | DYNR | 3653 [-11;3769] | 30.5 | 34.1 | 1533 [-3;1804] | 42.8 | 28.1 | 3728 [-27;4795] | 9.6 | 31 | 0.00 | 0.00 | |
| | OPENMX | -8 [-12;-3] | 72.2 | 77.5 | 0 [-5;6] | 96.0 | 0 | -17 [-31;-1] | 25.0 | 73.8 | 0.46 | 0.46 | |
| | GLLA | 1 [-4;7] | 97.4 | 92.5 | 6 [0;14] | 77.1 | 0 | -5 [-21;11] | 52.3 | 92 | 0.99 | 0.45 | 7 |
| | FDA | -0 [-5;5] | 98.9 | 92.3 | 3 [-3;10] | 89.2 | 0 | -4 [-20;13] | 53.8 | 92.5 | 0.98 | 0.45 | 0.55 |
| 1: Reference condition: STN 30% | DYNR | 2 [-12;2672] | 35.1 | 44.1 | 10 [-4;843] | 49.5 | 0.1 | 12 [-33;3796] | 13.9 | 40.9 | 0.37 | 0.37 | |
| | OPENMX | -7 [-15;0] | 66.4 | 84.1 | 0 [-5;8] | 93.6 | 0 | -16 [-38;5] | 28.5 | 79.5 | 0.45 | 0.45 | |
| | NLME | 2194 [383;3846] | 0.2 | 0.9 | 160 [-26;516] | 0.4 | 0 | 1749 [384;2747] | 0.4 | 1.4 | 0.50 | 0.10 | |
| | GLLA | 8 [-1;17] | 63.6 | 81.1 | 17 [9;25] | 14.0 | 0 | -7 [-31;17] | 37.1 | 91.8 | 0.79 | 0.45 | 15 |
| | FDA | -1 [-8;7] | 91.9 | 93.5 | 5 [-2;12] | 81.7 | 0 | -8 [-28;13] | 41.2 | 91.8 | 0.81 | 0.45 | 0.73 |
| 14: STN = 50% | DYNR | 1 [-16;2283] | 33.7 | 51.4 | 7 [-5;511] | 51.6 | 0 | 7 [-37;3647] | 16.3 | 51.4 | 0.39 | 0.39 | |
| | OPENMX | -7 [-18;4] | 58.9 | 89.8 | 1 [-7;9] | 86.6 | 0 | -15 [-43;12] | 27.8 | 87.5 | 0.45 | 0.45 | |
| | GLLA | -0 [-12;14] | 65.9 | 92.1 | 18 [8;29] | 15.6 | 0 | -24 [-57;8] | 19.9 | 78.8 | 0.00 | 0.43 | 16 |
| | FDA | 8 [-4;19] | 58.4 | 85.3 | 7 [-3;17] | 66.3 | 0 | 10 [-15;38] | 34.7 | 92.8 | 0.00 | 0.45 | 0.82 |

Table 7: relative bias (bias), percentage of estimations with an absolute relative bias below 10% (N10) and coverage (cov) of the estimated coefficients (decay rate $\gamma$, parameters associated with the gain $K\gamma$ and with the equilibrium value $y_{eq}\gamma$) for the different estimation methods and for different individual and measurement error (13: 10% STN, 1: 30% STN, 14: 50% STN). The median $R^2$ calculated with the estimated curve from the individual coefficients (R2r) and from the fixed effect coefficient (R2g) and the simulation signal without noise, as well as the embedding number or smoothing parameter used (d) are displayed.

Both one-step method base on state-space modeling seem to be robust to the variation of measurement error, whereas the two-step methods are strongly dependent on it (see Table 7). For low noises (simulation condition 13), all methods yield acceptable results. For a measurement noise of 50% of the maximum signal amplitude (simulation condition 14), the bias produced by the three two-step methods increases. In this situation, the FDA method seems to perform better.

Note that the optimal embedding dimension or smoothing parameter is lower for a lower noise and increases with the measurement error.

*Effect of other regression methods in the second step of two-step methods*

| params | | $\gamma$ | | | $K\gamma$ | | | $y_{eq}\gamma$ | | | R2 | | d |
|---|---|---|---|---|---|---|---|---|---|---|---|---|---|
| | **Method** | **bias** | **N10** | **cov** | **bias** | **N10** | **cov** | **bias** | **N10** | **cov** | **R2r** | **R2g** | **d** |
| 11: Reference condition | DYNR | 2 [-12;2672] | 35.1 | 44.1 | 10 [-4;843] | 49.5 | 0.1 | 12 [-33;3796] | 13.9 | 40.9 | 0.37 | 0.37 | |
| | OPENMX | -7 [-15;0] | 66.4 | 84.1 | 0 [-5;8] | 93.6 | 0 | -16 [-38;5] | 28.5 | 79.5 | 0.45 | 0.45 | |
| | NLME | 2194 [383;3846] | 0.2 | 0.9 | 160 [-26;516] | 0.4 | 0 | 1749 [384;2747] | 0.4 | 1.4 | 0.50 | 0.10 | |
| | GLLA | 8 [-1;17] | 63.6 | 81.1 | 17 [9;25] | 14.0 | 0 | -7 [-31;17] | 37.1 | 91.8 | 0.79 | 0.45 | 15 |
| | FDA | -1 [-8;7] | 91.9 | 93.5 | 5 [-2;12] | 81.7 | 0 | -8 [-28;13] | 41.2 | 91.8 | 0.81 | 0.45 | 0.73 |
| 16: Lm regressio | GLLA | -60 [-66;-53] | 0.0 | 0 | -8 [-14;-1] | 63.7 | 0 | -113 [-127;-97] | 0.0 | 0 | 0.00 | 0.00 | 5 |
| | FDA | -61 [-68;-53] | 0.0 | 0 | -38 [-43;-33] | 0.0 | 0 | -62 [-76;-47] | 0.0 | 0 | 0.38 | 0.38 | 0.91 |
| 17: Gee regressi | GLLA | -2 [-10;6] | 86.9 | 89.3 | 12 [5;19] | 38.8 | 0 | -25 [-46;-2] | 18.5 | 66 | 0.44 | 0.44 | 15 |
| | FDA | 0 [-6;7] | 93.6 | 92.9 | -3 [-9;4] | 92.7 | 0 | 6 [-14;26] | 44.3 | 92.2 | 0.45 | 0.45 | 0.82 |
| **params** | | **gamma** | | | **A** | | | **yogamma** | | | **R2** | | **d** |
| | **Method** | **bias** | **N10** | **cov** | **bias** | **N10** | **cov** | **bias** | **N10** | **cov** | **R2r** | **R2g** | **d** |
| 1 | DYNR | 2 [-12;2672] | 35.1 | 44.1 | 10 [-4;843] | 49.5 | 0.1 | 12 [-33;3796] | 13.9 | 40.9 | 0.37 | 0.37 | |
| | OPENMX | -7 [-15;0] | 66.4 | 84.1 | 0 [-5;8] | 93.6 | 0 | -16 [-38;5] | 28.5 | 79.5 | 0.45 | 0.45 | |
| | NLME | 2194 [383;3846] | 0.2 | 0.9 | 160 [-26;516] | 0.4 | 0 | 1749 [384;2747] | 0.4 | 1.4 | 0.50 | 0.10 | |
| | GLLA | 8 [-1;17] | 63.6 | 81.1 | 17 [9;25] | 14.0 | 0 | -7 [-31;17] | 37.1 | 91.8 | 0.79 | 0.45 | 15 |
| | FDA | -1 [-8;7] | 91.9 | 93.5 | 5 [-2;12] | 81.7 | 0 | -8 [-28;13] | 41.2 | 91.8 | 0.81 | 0.45 | 0.73 |
| 16 | GLLA | -60 [-66;-53] | 0.0 | 0 | -8 [-14;-1] | 63.7 | 0 | -113 [-127;-97] | 0.0 | 0 | 0.00 | 0.00 | 5 |
| | FDA | -61 [-68;-53] | 0.0 | 0 | -38 [-43;-33] | 0.0 | 0 | -62 [-76;-47] | 0.0 | 0 | 0.38 | 0.38 | 0.91 |
| 17 | GLLA | -2 [-10;6] | 86.9 | 89.3 | 12 [5;19] | 38.8 | 0 | -25 [-46;-2] | 18.5 | 66 | 0.44 | 0.44 | 15 |
| | FDA | 0 [-6;7] | 93.6 | 92.9 | -3 [-9;4] | 92.7 | 0 | 6 [-14;26] | 44.3 | 92.2 | 0.45 | 0.45 | 0.82 |

Table 8: relative bias (bias), percentage of estimations with an absolute relative bias below 10% (N10) and coverage (cov) of the estimated coefficients (decay rate $\gamma$, parameters associated with the gain $K\gamma$ and with the equilibrium value $y_{eq}\gamma$) with a two steps procedure for the different calculations of the derivative and for different regression methods (1: mixed effect linear model, 16: linear model, 17: generalized estimating equation with exchangeable covariance structure). The median $R^2$ calculated with the estimated curve from the individual coefficients (R2r) and from the fixed effect coefficient (R2g) and the simulation signal without noise, as well as the embedding number or smoothing parameter used (d) are displayed.

In order to illustrate the utility of the mixed effect regression, we performed the two-step analysis of the reference simulated panel data but using a standard linear regression model as a second step (simulation condition 16, see Table 8). Although the coefficients of the first order ODE used to generate the panel are distributed along a normal distribution, the regression without the random effects produced strongly biased results for all estimated parameters. The same procedure but with generalized estimating equations (simulation condition 17) leads to results similar to the mixed effect regression, but without the possibility to calculate individual coefficients.

## Application

We used the presented model to fit cardiological data obtained from effort tests of 30 Spanish amateur running athletes. Physiological characteristics and performances of these athletes are given in supplementary Table S2. We want to examine whether the proposed model is able to explain most of the variance observed in cardiological data, and the estimated coefficients produce cardiac indices sensible to cardio-respiratory condition among such homogeneous population.

The athletes performed graded exercise on a PowerJog J series treadmill connected to a CPX MedGraphics gas analyzer system (Medical Graphics, St Paul, MN, USA) with cycle-to-cycle measurements of respiratory parameters -including $VO_2$, and HR- with a 12 lead ECG (Mortara). The stress test consisted of an 8-10 min warm up period of 5 $km.h^{-1}$ followed by continuous 1$km.h^{1}$ by minute speed increase until the maximum effort was reached. Power developed during the effort test was calculated using the formula described by the American College of Sport Medicine (ACSM). The latter determines an approximate $VO_2$ of runners (Ferguson, 2014) associated to the Hawley and Noakes equation that links oxygen consumption to mechanical power (Hawley & Noakes, 1992). The resulting panel consisted in 30 recordings of Heart Rate (HR) and power spent during effort, each containing between 200 and 500 measurements with non-equidistant time-steps ranging from 1 to 5 s.

Most common analyzes of HR during effort test consist in the calculation of easily calculable indexes, such as the Heart Resting Rate (HRR; Cole, Blackstone, Pashkow, Snader, & Lauer, 1999) or the heart rate reserve (%HRR; Swain & Leutholtz, 1997). The HRR calculated here is the standard HRR60, which is the difference between the HR at the onset of the recovery and the HR 60 seconds later. The %HRR is the difference between the resting heart rate and the maximum heart rate reached during the effort test. These cardiac indices, considered among the best for aerobic fitness evaluation (Borresen & Lambert, 2008; Valentini & Parati, 2009), use only a small fraction of the information contained in the entire effort test. Recent approaches based on first order differential equations have shown the potential ability to model oxygen consumption (Artiga Gonzalez et al., 2017) or HR measurement (Zakynthinaki, 2015) using dynamical analysis. However, the complex estimation procedure of the 9 coefficient-multi differential equation model proposed by by Zakynthunaki and co-authors is a serious drawback. Knowing that the heart is mainly a self-regulated system responding to the energy consumption of the body (Valentini & Parati, 2009), we applied the multilevel two-step procedure presented to analyze the HR dynamics during effort tests. Correlation of the subsequent estimated decay time, equilibrium value and gain with the main indicators of athlete's performing capacities was also computed. These indicators are the maximal $VO_2$ reached during the exercise, the maximal speed reached, and the values of speed at the two Ventilatory Thresholds (VTs), corresponding to the lactic apparition (VT1) and the accumulation (VT2) threshold (Binder et al., 2008).

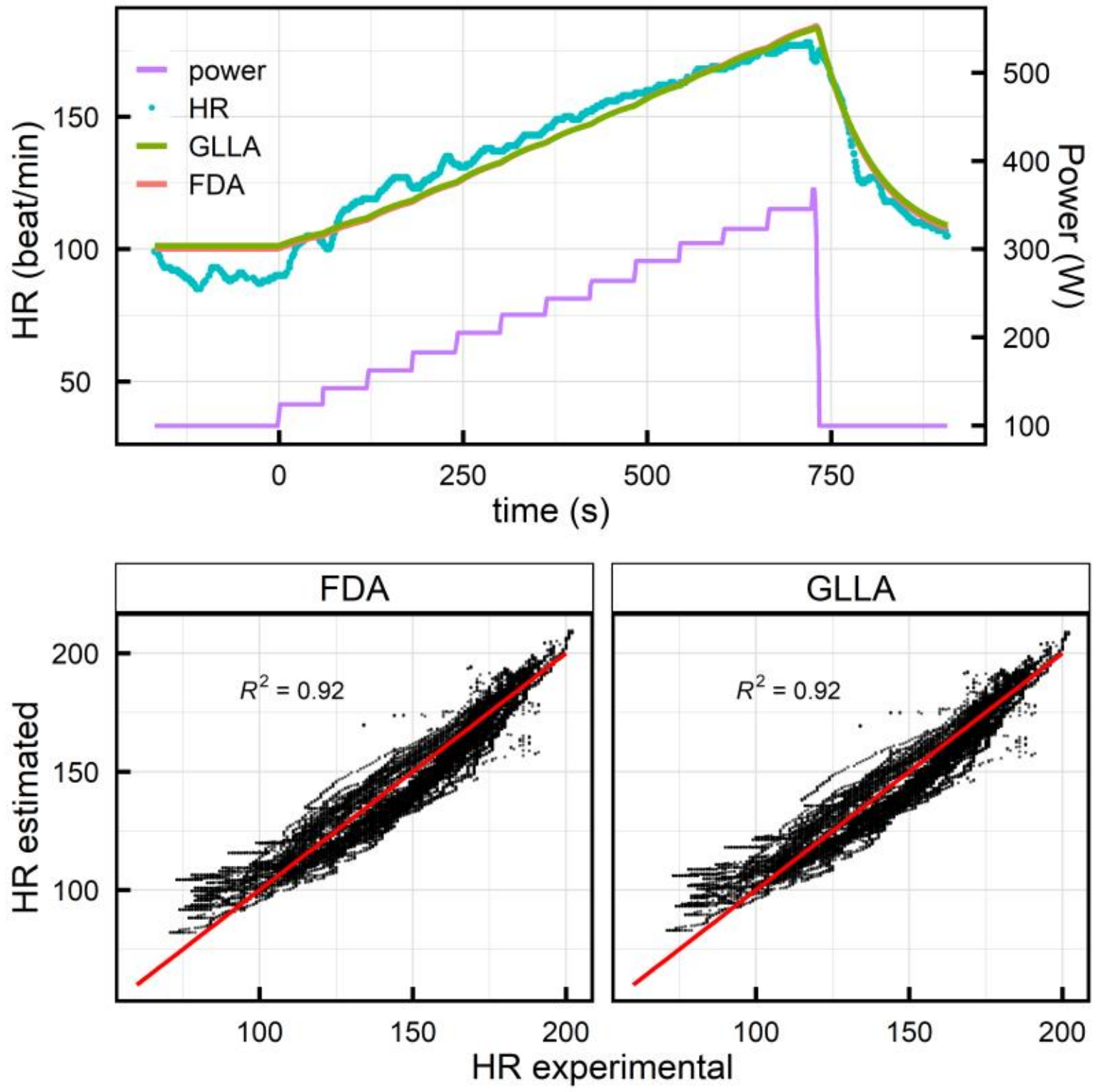

*Figure 5*: top: Adjustment with the first order differential equation regression of heart rate (HR) measurements performed during an effort test on treadmill. Bottom: totality of the estimated HR values compared to the true value, for the two methods considered.

An example of the adjustments obtained for the two-step methods considered is shown in Figure 5, together with a comparison between the estimated points and the actual measurements. *dynr* algorithm did not converge in this case, nor did *nlmeODE*'s. The use of *mxExpectationStateSpace* wasn't successful because of its necessity to have initial value close to the true ones.

The optimization procedure to determine the embedding dimension yielded an embedding dimension of 13 for GLLA, and a smoothing parameter of 0.56 for the FDA spline method. In line with our simulation study, the FDA and GLLA analysis reproduced extremely well the variation of heart rate during effort ($R^2$ = 0.92). For the rest of this section we will only consider the FDA results. These yielded a mean decay time of $107 \pm 4$ s, a gain of $0.41 \pm 0.05$ beat/min/W (meaning that an effort increase of 100 W increases the heart rate of 41 beat/min) and an equilibrium value of $103.7 \pm 7.7$ beat/min. The estimated equilibrium value was overestimated, as can be seen in Figure 6 (the estimated curves are above the experimental signal, and the low HR estimated values in the bottom plots deviate from the identity). The HR equilibrium value calculated on the experimental data yields a value of 90 beat/min.

The FDA+linear mixed effects regression method allowed to determine each individual's coefficients, thus giving the opportunity to study the link between these and other individual variables. Table 9 displays the correlation between the individual estimated HR decay time, gain and equilibrium value with HRR, %HRR, maximum oxygen consumption ($VO_2$ max), maximum speed achieved during the effort test (max speed) and the speed at each ventilatory threshold (VT1 and VT2).

| index | Gain K | Decay time $\tau$ | $HR_{eq}$ | HRR | %HRR |
|---|---|---|---|---|---|
| HRR | 0.10 | -0.70*** | -0.54** | | |
| %HRR | 0.34 | 0.18 | -0.57*** | 0.22 | |
| $VO_2$ max | -0.81*** | 0.14 | 0.06 | 0.03 | 0.08 |
| Max speed | -0.51** | -0.10 | -0.22 | 0.30 | 0.31 |
| VT1 | -0.09 | 0.14 | -0.37* | 0.13 | 0.29 |
| VT2 | -0.42* | -0.03 | -0.18 | 0.20 | 0.19 |

Table 9: Spearman correlation index between individual dynamical coefficient estimated (the gain $K$, the decay time $\tau$ and the equilibrium value $HR_{eq}$) and standard cardiovascular indices: Heart resting rate (HRR), heart rate reserve (%HRR), maximum oxygen consumption (VO2 max), maximum speed achieved during the test (Max speed), and the speed at the ventilatory thresholds (VT1 and VT2). Significance is indicated as follows: *: $p<0.05$; **: $p < 0.01$, ***: $p < 0.001$

A clear difference between our dynamical analysis and the standard cardiac indices can be observed. On one hand, the HRR and %HRR do not present any correlations with the other performance indices ($VO_2$ max, VT1, VT2, maximum speed). On the other hand, the cardiac gain (i.e. the proportionality between the energy expenditure increase and the steady state heart rate increase it will cause) is strongly inversely correlated with VO2 max, inversely correlated with the maximum speed reached during the test and slightly inversely correlated with the second ventilatory threshold. This means that a lower heart rate increase for a given effort is associated with better physical and anaerobic performances. The estimated decay time is inversely correlated with HRR, consistent with the fact that a faster dynamic will lead to higher HRR. The estimated HR equilibrium value is inversely correlated to HRR and %HRR (see Table 9).

**Discussion**

Longitudinal data measuring a variable experiencing multiple excitations and return to equilibrium phases are common in various fields such as behavioral science, where a large number of psychological variables can be influenced by repeated external life events, such as cognitive abilities by sex hormones (Courvoisier et al., 2013), and medicine, where disease activity is repeatedly influenced by exogeneous variable such as drug intakes. The work of Trail and co-authors (Trail et al., 2014) showed that considering explicit excitation processes of the dynamics observed can be of high interest for behavioral science. To further promote the use of such a method, it is essential to provide its estimator characteristics and its condition of use.

We prove here that the use of a two-step estimation procedure using a FDA method to calculate the derivative and a linear mixed regression to estimate the coefficients of a first order differential equation that includes an excitation term, is an accurate and robust method to reproduce noisy outcomes following self-regulated processes. Such a method accounts for inter-individual differences and allows the extraction of three parameters that characterize the dynamics: the equilibrium value, the outcome's characteristic time (decay time) and the response amplitude to the perturbation (gain).

The simulations performed show that this method produces unbiased parameters when having at least 5 points per decay time, more than 30 observations per individual, a measurement noise up to 50% of the maximum signal amplitude and an excitation process lasting several time points. Not taking the excitation into account can lead to severe bias and deprive the researcher from the information provided by the gain and its effect on the dynamics.

A ready to use program implementing this model has been developed and is proposed as a full R package in CRAN containing the analyses and simulation features (Mongin et al., 2019).

The simulation results highlight an already known intrinsic limitation of the two-step method: the calculation of the derivative is a source of bias. As the methods discussed to estimate the derivative are a local approximation of the real derivative, the estimation is biased when the observed variable does not have enough measurements near curvature changes i.e. when the derivative varies quickly. The bias in the estimation of the derivative introduce bias in the estimation of the coefficients when performing the regression. This is the case when the decay time is small, i.e. when there is an important curvature change between two observed points. This is also the case when the signal analyzed is generated by short excitations (i.e. when the data contains only one observation during the excitation process). Indeed, during a punctual excitation, the derivative presents a discontinuity at the onset of the excitation (i.e. the variable observed rises instantaneously), and a change of sign at its end (i.e. when the excitation ends, the variable starts to decrease). The local approximation performed when calculating the derivative cannot account for such a brief variation of the derivative. This bias is reduced when the excitation lasts for several measurements, as observed in the simulation study.

Two methods to calculate the derivative within the two-step estimation procedure have been compared. In most of the simulation conditions, the FDA method outperforms the GLLA. Note that we did not try, as proposed in other study, to use GLLA at higher order (Chow, Bendezú, et al., 2016). As these two methods are a linear approximation of the true derivative, which is an exponential behavior, they tend to induce an underestimation of the decay rate. The FDA method ability to reproduce curvature leads to better derivative estimates, in line with the results of a study comparing these methods on non-linear differential equations (Chow, Bendezú, et al., 2016). The two-step procedure using generalized estimating equations provides similar fixed effect estimates than the procedure using a mixed effects model but hinder the access to individual dynamic coefficients. It should be considered when computational speed is a concern.

The two-step procedure without random effects included in the second step led to important bias in the estimation of all parameters, thus highlighting the importance of considering random effects for this estimation method.

The ASDE method cannot really be applied using the analytical solution of the differential equation. Indeed, when the excitation can have any shape, the analytical solution would be tedious or impossible to derive analytically. The use of numerical solutions instead allows to use this method in the case of multiple excitations, but still led to two important caveats in the context of this study. First, the use of non-linear regression with random effects in a panel of 50 individuals and 50 observations each resulted in a computing time of at least 12 hours for a single estimation on a standard computer. Secondly, compared with the two-step approach, this approach needs the estimation of a supplementary parameters: the initial value of the observed variable. As shown previously, initial values that are far from the real value can lead to strongly biased estimates (Hu & Huang, 2018) as the optimization algorithms can find local minimums. In our case, even though initial values were close to the true values, this method resulted in strongly biased parameters and overestimated p values.

The single-step estimation of a state-space model embedded in *mxExpectationStateSpace* of the *OpenMx* package led to proper estimation of the fixed effect of the differential equation coefficients in most of the simulation conditions. When the system experiences short excitations, it is the only method yielding acceptable estimates. Given the necessity to provide initial values for the estimated parameters and the absence of random effects included in *mxExpectationStateSpace*, and given the good results provided by the two-step method under the conditions described before, its use is of interest for strongly noisy signals or for variables experiencing short excitation processes.

Regarding the single-step procedure included in the *dynr* package, the authors didn't manage to include random effects for the decay time and the equilibrium value in the model, although theoretically possible through the introduction of artificial latent variables. For the reference simulation condition, although producing proper median estimate, this procedure tends to produce strongly overestimated parameters and underestimated standard errors. The strongly biased results this method yields when changing the parameters (to smaller damping rates, or shorter excitations, or null equilibrium values) are caused by a greater sensibility to starting values than *mxExpectationStateSpace.* Indeed, the results provided by *dynr* are biased when the range of the starting values provided to *dynr* in our simulation are too far from the real ones.

The simulation results obtained with *dynr* should not be taken as a claim that this tool is not appropriate for the present problem. There may be ways to specify the model that reduce its sensibility to starting values, and *dynr* provides a broad range of tools and estimation methods that we did not explore. Among them, the discrete time state space model and its estimation through the Kalman or Kim filter (Chow, Ho, Hamaker, & Dolan, 2010), and the implementation of random effect as supplementary latent variables could improve or change the results.

The application of the two-step estimation method in a sport cardiology setting showed promising results to explain behavior in real acute stressful conditions. The three parameters obtained greatly summarized and explained complex dynamics, as shown by the $R^2$ of 0.91. The estimated gain is a key result of such analysis as it clearly shows the response of the system (the heart) to the excitation (the effort). It is sensible to cardiovascular performances indices in a homogeneous population of athletes, where standard cardiac indices are not, and indicates that better cardiovascular performances are linked with a smaller heart rate increase for a given effort, in line with established results (Bellenger et al., 2016; Valentini & Parati, 2009). However, our method yielded a biased estimate of the equilibrium value. This is because during effort tests, the heart rate is known to return to a higher equilibrium value after effort. It then decreases back to its resting value on a longer time scale due to several mechanisms such as the reduction of blood volume (i.e. dehydration) or the evacuation of the heat accumulated during the muscular contractions (Lambert, Mbambo, & Gibson, 1998; Savin, Davidson, & Haskell, 1982; Wyss, Brengelmann, Johnson, Rowell, & Niederberger, 1974). This change in the equilibrium value (allostasis) is not included in the present state of our model, that assumes that the equilibrium value is identical before and after the effort.

Other applications are numerous, and range from the bio-medical fields, where response to drugs could be modeled in a similar fashion, to the behavioral field where any measurement of self-regulated processes such as mood matching the requirements described above (30 observations, 5 measurement per typical decay time and an excitation process lasting several observations) could be of interest (Trail et al., 2014).

There are limitations to our study and model.

First, our model supposes constant coefficients (decay time, gain and equilibrium value) during the whole dynamics. It cannot in its present state model regime switching phenomena. This active domain of research has been the subject of considerable developments based on state space modeling (Chow et al., 2018; Chow & Zhang, 2013) and of recent advances based on the SEM framework (Boker, 2015; McKee, Rappaport, Boker, Moskowitz, & Neale, 2018).

Secondly, although proving the value of the method presented, our simulation study considered a restricted amount of estimation techniques. It would be of great interest to test other estimation methods on a first order differential equation including an excitation term. The estimation of the p values for any excitation process will require some further study, as shown by the coverage, and the development of a variance inflation factor to correct for the two-step process. Finally, extensions of the model to relax some of the assumptions (e.g., similar equilibrium value before and after effort test) and integrate a variation of the parameters over time could be of great interest.

In conclusion, the dynamic model using a two-step method using an FDA spline approach to calculate the derivative and a mixed effect regression for the estimation of the coefficients of a first order differential equation provided accurate results in a simulation study showing a large panel of conditions. Together with the interesting findings of its application on cardiological data recorded during effort test, it opens the way to new analyses in numerous fields studying self-regulated processes.

Acknowledgment

The authors would like to thank the three anonymous reviewers and associate editor Dr. Zhang for their helpful and constructive comments.

## Supplementary tables

| | params | $\gamma$ | | | $K\gamma$ | | | $y_{eq}\gamma$ | | | $R^2$ | | d |
|---|---|---|---|---|---|---|---|---|---|---|---|---|---|
| | Method | bias | N10 | cov | bias | N10 | cov | bias | N10 | cov | R2r | R2g | d |
| 1: Reference condition | DYNR | 2 [-12;2672] | 35.1 | 44.1 | 10 [-4;843] | 49.5 | 0.1 | 12 [-33;3796] | 13.9 | 40.9 | 0.37 | 0.37 | |
| | OPENMX | -7 [-15;0] | 66.4 | 84.1 | 0 [-5;8] | 93.6 | 0 | -16 [-38;5] | 28.5 | 79.5 | 0.45 | 0.45 | |
| | NLME | 2194 [383;3846] | 0.2 | 0.9 | 160 [-26;516] | 0.4 | 0 | 1749 [384;2747] | 0.4 | 1.4 | 0.50 | 0.10 | |
| | GLLA | 8 [-1;17] | 63.6 | 81.1 | 17 [9;25] | 14.0 | 0 | -7 [-31;17] | 37.1 | 91.8 | 0.79 | 0.45 | 15 |
| | FDAFDA | -1 [-8;7] | 91.9 | 93.5 | 5 [-2;12] | 81.7 | 0 | -8 [-28;13] | 41.2 | 91.8 | 0.81 | 0.45 | 0.73 |
| 2: $\gamma$ = 1/5 | DYNR | -8 [-16;720] | 20.6 | 16.1 | -4 [-13;437] | 40.9 | 27.9 | -5 [-29;996] | 17.6 | 30.6 | 0.61 | 0.61 | |
| | OPENMX | -12 [-18;-6] | 34.4 | 25.6 | -8 [-14;-2] | 65.6 | 0 | -17 [-33;2] | 27.2 | 47.8 | 0.66 | 0.66 | |
| | GLLA | -3 [-8;4] | 94.1 | 87.3 | 4 [-2;12] | 80.9 | 0 | -12 [-29;7] | 39.6 | 83.3 | 0.96 | 0.65 | 11 |
| | FDAFDA | -10 [-16;-4] | 46.3 | 38 | -5 [-11;3] | 84.6 | 0 | -17 [-33;0] | 28.8 | 71.9 | 0.94 | 0.65 | 0.73 |
| 3: $\gamma$ =1/10 | DYNR | -2 [-13;1728] | 36.9 | 38.1 | 4 [-6;736] | 53.0 | 0.8 | 1 [-29;2387] | 15.3 | 39.7 | 0.48 | 0.48 | |
| | OPENMX | -8 [-14;-1] | 63.7 | 64.4 | -2 [-8;5] | 92.2 | 0 | -16 [-33;6] | 26.5 | 66.6 | 0.53 | 0.53 | |
| | GLLA | -2 [-9;6] | 92.0 | 90.4 | 9 [1;16] | 59.6 | 0 | -14 [-36;6] | 31.5 | 77.7 | 0.93 | 0.54 | 11 |
| | FDAFDA | -4 [-10;3] | 90.1 | 86.4 | 3 [-3;10] | 89.7 | 0 | -12 [-31;9] | 35.8 | 83.9 | 0.89 | 0.54 | 0.73 |
| 4: $\gamma$ = 1/20 | DYNR | 3108 [-14;3649] | 25.0 | 40.4 | 484 [-2;911] | 40.9 | 0 | 4105 [-32;5200] | 10.3 | 39.9 | 0.01 | 0.01 | |
| | OPENMX | -8 [-16;1] | 60.8 | 92 | 2 [-5;9] | 91.6 | 0 | -19 [-40;6] | 24.0 | 86.8 | 0.38 | 0.38 | |
| | GLLA | 5 [-4;15] | 72.8 | 88.3 | 17 [9;25] | 12.9 | 0 | -13 [-38;14] | 29.9 | 86.7 | 0.46 | 0.38 | 15 |
| | FDAFDA | 0 [-9;9] | 87.5 | 93.9 | 6 [-2;12] | 78.4 | 0 | -6 [-29;19] | 37.1 | 91.8 | 0.63 | 0.39 | 0.73 |
| 5: Excitation: one step | DYNR | 2580 [-31;2804] | 1.7 | 10.1 | 392 [-3;504] | 25.8 | 0.1 | 2965 [-44;3329] | 1.2 | 9.5 | 0.02 | 0.02 | |
| | OPENMX | -14 [-22;-5] | 28.3 | 88.4 | -0 [-5;6] | 97.0 | 0 | -20 [-32;-6] | 17.4 | 85.3 | 0.30 | 0.30 | |
| | GLLA | -4 [-14;7] | 70.6 | 87.5 | 14 [8;23] | 20.8 | 0 | -12 [-27;5] | 39.3 | 78.7 | 0.79 | 0.29 | 11 |
| | FDA | -11 [-20;-1] | 40.7 | 58.1 | 2 [-4;10] | 90.3 | 0 | -16 [-30;-1] | 28.3 | 65.7 | 0.50 | 0.30 | 0.64 |
| | DYNRFDA | 2650 [-176;2855] | 3.8 | 1.6 | 37 [-29;84] | 9.3 | 0 | 2683 [-193;3095] | 3.9 | 1.6 | 0.00 | 0.00 | |
| 6: Excitation: 3 points | OPENMX | -9 [-17;-0] | 58.4 | 99.8 | -14 [-20;-8] | 20.6 | 0 | -8 [-18;3] | 58.2 | 99.8 | 0.02 | 0.02 | |
| | GLLA | -11 [-25;1] | 43.5 | 70.8 | -73 [-78;-70] | 0.0 | 0 | -2 [-19;14] | 55.8 | 86.8 | 0.80 | 0.00 | 25 |
| | FDA | -26 [-36;-16] | 2.4 | 7.2 | -3 [-12;5] | 79.9 | 0 | -26 [-38;-15] | 4.8 | 18.1 | 0.81 | 0.02 | 0.73 |
| | DYNRFDA | 2558 [-128;2694] | 4.4 | 4.1 | -47 [-68;45] | 5.4 | 0 | 2764 [-142;3124] | 3.1 | 4.1 | 0.00 | 0.00 | |
| 7: Ex | OPENMX | 4 [-7;15] | 72.3 | 100 | -24 [-31;-17] | 0.6 | 0 | 9 [-5;24] | 48.9 | 99.9 | 0.02 | 0.02 | |

| | params | $\gamma$ | | | $K\gamma$ | | | $y_{eq}\gamma$ | | | $R^2$ | | d |
|---|---|---|---|---|---|---|---|---|---|---|---|---|---|
| | Method | bias | N10 | cov | bias | N10 | cov | bias | N10 | cov | R2r | R2g | d |
| | GLLA | -17 [-51;-3] | 26.0 | 53.1 | -96 [-99;-93] | 0.0 | 0 | -1 [-43;18] | 47.3 | 82.8 | 0.28 | 0.00 | 25 |
| | FDA | -37 [-97;-18] | 3.4 | 15.3 | -13 [-20;-4] | 33.7 | 0 | -40 [-107;-15] | 5.1 | 25 | 0.00 | 0.01 | 0.55 |
| | DYNRFDA | 2519 [-236;2646] | 1.1 | 2.3 | -75 [-100;18] | 7.1 | 0 | 3062 [-306;3380] | 1.4 | 2.2 | 0.00 | 0.00 | |
| 8: Excitation: 10 points | OPENMX | 11 [-1;23] | 46.0 | 99.7 | -44 [-53;-35] | 0.0 | 0 | 30 [13;47] | 6.9 | 99.5 | 0.03 | 0.03 | |
| | GLLA | -28 [-40;-15] | 4.5 | 10.7 | -73 [-77;-70] | 0.0 | 0 | -9 [-25;9] | 46.1 | 79.2 | 0.91 | 0.00 | 15 |
| | FDA | -8 [-21;4] | 53.8 | 79.5 | -99 [-112;-86] | 0.0 | 0 | 26 [8;47] | 11.5 | 53.2 | 0.00 | 0.00 | 0.64 |
| | DYNRFDA | 2260 [-112;2338] | 6.1 | 33.9 | 851 [-47;960] | 12.7 | 0 | 1693 [-96;2313] | 6.3 | 33.7 | 0.42 | 0.42 | |
| 9: 30 Observations | OPENMX | -1 [-45;49] | 21.7 | 98.2 | -0 [-19;23] | 44.6 | 0 | -7 [-45;39] | 24.1 | 97.8 | 0.45 | 0.45 | |
| | GLLA | -106 [-116;-95] | 0.0 | 0 | -34 [-43;-25] | 0.0 | 0 | -101 [-116;-85] | 0.0 | 0 | 0.00 | 0.00 | 23 |
| | FDA | -99 [-118;-80] | 0.0 | 0.2 | -43 [-57;-28] | 0.3 | 0 | -87 [-118;-55] | 0.0 | 70.3 | 0.00 | 0.00 | 0.18 |
| | DYNRFDA | 7 [-16;2752] | 28.4 | 49 | 14 [-6;921] | 41.7 | 3.3 | 23 [-38;4002] | 12.9 | 46.9 | 0.34 | 0.34 | |
| 10: 20 individuals | OPENMX | -7 [-18;5] | 57.4 | 92.7 | 0 [-9;13] | 76.7 | 0.1 | -16 [-47;16] | 23.6 | 88 | 0.46 | 0.46 | |
| | GLLA | -4 [-17;10] | 62.4 | 88.1 | 10 [-2;22] | 49.4 | 0 | -21 [-56;16] | 19.8 | 80.4 | 0.89 | 0.45 | 11 |
| | FDA | -2 [-12;11] | 69.7 | 92.3 | 5 [-6;16] | 69.8 | 0 | -8 [-43;28] | 27.3 | 89.6 | 0.84 | 0.46 | 0.73 |
| | DYNRFDA | 2282 [-11;2658] | 32.9 | 32.7 | 473 [-2;828] | 47.2 | 0 | 3047 [-28;3766] | 9.5 | 30.5 | 0.07 | 0.07 | |
| 11: 100 individuals | OPENMX | -7 [-13;-1] | 72.1 | 71.5 | 1 [-4;6] | 96.9 | 0 | -16 [-30;-1] | 24.0 | 70 | 0.45 | 0.45 | |
| | GLLA | 7 [1;14] | 68.8 | 64.5 | 17 [11;23] | 5.2 | 0 | -7 [-24;10] | 50.1 | 88.8 | 0.74 | 0.45 | 15 |
| | FDAFDA | -1 [-6;5] | 97.4 | 92 | 5 [0;10] | 88.8 | 0 | -7 [-22;8] | 53.7 | 88.7 | 0.79 | 0.45 | 0.73 |
| 12: $Y_{eq} = 0$ | DYNR | 2281 [-13;2676] | 27.0 | 35.9 | 472 [-3;851] | 44.1 | 0.1 | 0.00 [-0.00;0.00] | 99.5 | 36.1 | 0.07 | 0.07 | |
| | OPENMX | -2 [-10;7] | 82.5 | 93.7 | -2 [-9;5] | 89.0 | 0 | 0.00 [0.00;0.00] | 100.0 | 97.2 | 0.45 | 0.45 | |
| | GLLA | 2 [-6;11] | 83.2 | 92.6 | 14 [6;22] | 27.3 | 0 | -0.00 [-0.00;0.00] | 100.0 | 76.9 | 0.84 | 0.45 | 13 |
| | FDAFDA | -1 [-8;7] | 91.2 | 93.8 | 5 [-2;12] | 81.1 | 0 | -0.00 [-0.00;0.00] | 100.0 | 89.7 | 0.81 | 0.45 | 0.73 |
| 13: STN = 10% | DYNR | 3653 [-11;3769] | 30.5 | 34.1 | 1533 [-3;1804] | 42.8 | 28.1 | 3728 [-27;4795] | 9.6 | 31 | 0.00 | 0.00 | |
| | OPENMX | -8 [-12;-3] | 72.2 | 77.5 | 0 [-5;6] | 96.0 | 0 | -17 [-31;-1] | 25.0 | 73.8 | 0.46 | 0.46 | |
| | GLLA | 1 [-4;7] | 97.4 | 92.5 | 6 [0;14] | 77.1 | 0 | -5 [-21;11] | 52.3 | 92 | 0.99 | 0.45 | 7 |
| | FDAFDA | -0 [-5;5] | 98.9 | 92.3 | 3 [-3;10] | 89.2 | 0 | -4 [-20;13] | 53.8 | 92.5 | 0.98 | 0.45 | 0.55 |
| 14 : | DYNR | 1 [-16;2283] | 33.7 | 51.4 | 7 [-5;511] | 51.6 | 0 | 7 [-37;3647] | 16.3 | 51.4 | 0.39 | 0.39 | |

| | params | $\gamma$ | | | $K\gamma$ | | | $y_{eq}\gamma$ | | | $R^2$ | | d |
|---|---|---|---|---|---|---|---|---|---|---|---|---|---|
| | **Method** | **bias** | **N10** | **cov** | **bias** | **N10** | **cov** | **bias** | **N10** | **cov** | **R2r** | **R2g** | **d** |
| | OPENMX | -7 [-18;4] | 58.9 | 89.8 | 1 [-7;9] | 86.6 | 0 | -15 [-43;12] | 27.8 | 87.5 | 0.45 | 0.45 | |
| | GLLA | -0 [-12;14] | 65.9 | 92.1 | 18 [8;29] | 15.6 | 0 | -24 [-57;8] | 19.9 | 78.8 | 0.00 | 0.43 | 15 |
| | FDAFDA | 8 [-4;19] | 58.4 | 85.3 | 7 [-3;17] | 66.3 | 0 | 10 [-15;38] | 34.7 | 92.8 | 0.00 | 0.45 | 0.82 |
| 15: Lm regressio | DYNR | 43 [-250;57] | 0.1 | 0.1 | NA [NA;NA] | NA | 0 | 79 [50;97] | 0.0 | 0 | 0.00 | 0.00 | |
| | OPENMX FDA | 8 [2;13] | 71.0 | 98.5 | NA [NA;NA] | NA | 0 | 179 [160;202] | 0.0 | 0 | NA | NA | |
| 16: Gee regressi | GLLA | -78 [-82;-72] | 0.0 | 0 | NA [NA;NA] | NA | 0 | -11 [-24;5] | 43.6 | 70.9 | 0.00 | 0.00 | 13 |
| | FDAFDA | -81 [-86;-76] | 0.0 | 0 | NA [NA;NA] | NA | 0 | -38 [-49;-24] | 0.6 | 2.5 | 0.00 | 0.00 | 0.82 |
| 17: Homogeneous equation | GLLA | -60 [-66;-53] | 0.0 | 0 | -8 [-14;-1] | 63.7 | 0 | -113 [-127;-97] | 0.0 | 0 | 0.00 | 0.00 | 5 |
| | FDA | -61 [-68;-53] | 0.0 | 0 | -38 [-43;-33] | 0.0 | 0 | -62 [-76;-47] | 0.0 | 0 | 0.38 | 0.38 | 0.91 |
| | GLLA | -2 [-10;6] | 86.9 | 89.3 | 12 [5;19] | 38.8 | 0 | -25 [-46;-2] | 18.5 | 66 | 0.44 | 0.44 | 15 |
| | FDAFDA | 0 [-6;7] | 93.6 | 92.9 | -3 [-9;4] | 92.7 | 0 | 6 [-14;26] | 44.3 | 92.2 | 0.45 | 0.45 | 0.82 |

Table S1: summary of the simulation results. For each simulation condition (left column, see Table 1) and method, the median [2.5;97.5%] relative bias of the estimated parameter, the percentage of estimation with absolute relative bias below 10% and the coverage of the three parameters (decay rate $\gamma$, parameters associated with the gain $K\gamma$ and with the equilibrium value $y_{eq}\gamma$) are given. The median $R^2$ calculated with the estimated curve from the individual coefficients (R2r) and from the fixed effect coefficient (R2g) and the simulation signal without noise, as well as (d) the embedding dimension (for GLLA) or smoothing parameter (for FDA spline), are provided. Light grey cells are condition reaching at least 50% of the estimated parameter below 10% of relative bias. The stronger grey indicate situation were the estimation explain more than 80% of the initial simulated parameter (R2 random above 0.8).

| | |
|---|---|
| Subject number | 30 |
| Age (year) | 36.07 (8.19) |
| Height (cm) | 174.26 (6.82) |
| Weight (kg) | 72.52 (9.70) |
| BMI | 23.83 (2.39) |
| Sex = male (%) | 28 (93.3) |
| Fat % | 16.15 (2.65) |
| $VO_2$ max (mL/min/kg) | 50.18 (6.16) |
| Maximum Speed | 17.71 (2.07) |
| Maximum HR | 182.33 (7.83) |
| VT1 speed (km/h) | 9.23 (1.52) |
| VT2 (km/h) | 13.20 (1.88) |
| HRR (beat/min) | 36.07 (8.91) |

Table S2: population characteristics of the 30 amateur athletes considered. Performance characteristics are the maximum dioxygen consumption (VO2 max), the maximum speed achieved during the effort test, the maximum heart rate, and the speed corresponding to the first ventilatory threshold (VT1) and the second one (VT2)

**Citations and References**